\documentclass{aa}  
\usepackage{graphicx}
\usepackage{multirow}
\newcommand{\bm}[1]{{\mbox{\boldmath $#1$}}}
\usepackage{color}
\usepackage{here}

\usepackage[colorlinks=true,citecolor=blue]{hyperref}
\usepackage{arydshln}
\usepackage{natbib}

\usepackage{booktabs}

\usepackage{ulem}

\usepackage{txfonts}
\begin{document}

   \title{
   Three-dimensional non-kinematic simulation of \\
   post-emergence evolution of bipolar magnetic regions \\
   and Babcock-Leighton dynamo of the Sun
   }
   
   \titlerunning{3D non-kinematic simulation of the BL solar dynamo}

   \author{Yuto Bekki
          \and
          Robert H. Cameron
          }

   \institute{Max-Planck-Institut f{\"u}r Sonnensystemforschung,
              Justus-von-Liebig-Weg 3, 37077 G{\"o}ttingen, Germany\\
            \email{\href{mailto:bekki@mps.mpg.de}{bekki@mps.mpg.de}}
             }

   \date{Received <-->; accepted <-->}

 
  \abstract
   {The Babcock-Leighton flux-transport model is a widely-accepted dynamo model of the Sun that can explain many observational aspects of the solar magnetic activity.
   This dynamo model has been extensively studied in a two-dimensional (2D) mean-field framework in both kinematic and non-kinematic regimes.
   Recent three-dimensional (3D) models have been restricted to the kinematic regime.
   In these models, the surface poloidal flux is produced by the emergence of bipolar magnetic regions (BMRs) that are tilted according to Joy's law.
   }
   {
   We investigate the prescription for emergence of a BMR in 3D non-kinematic simulations.
   In particular, we examine the effect of the radial extent of the BMR.
   We also report initial results of cyclic Babcock-Leighton dynamo simulation.
   }
   {
   We extend a conventional 2D mean-field model of the Babcock-Leighton flux-transport dynamo into 3D non-kinematic regime, in which a full set of magnetohydrodynamic (MHD) equations are solved in a spherical shell using a Yin-Yang grid.
   The large-scale mean flows  such as differential rotation and meridional circulation are driven not by rotationally-constrained convection but by the parameterized $\Lambda$-effect in this model. 
   For the induction equation, we use a Babcock-Leighton $\alpha$-effect source term by which the surface BMRs are produced in response to the dynamo-generated toroidal field inside the convection zone. 
    }
   {
   We find that, in the 3D non-kinematic regime, the tilt angle of a newly-emerged BMR is very sensitive to the prescription for the subsurface structure of the BMR (particularly its radial extent). 
   Anti-Joy tilt angles are found unless the BMR is deeply embedded in the convection zone.
   We also find that the leading spot tends to become stronger (higher field strengths) than the following spot.
   The anti-Joy's law trend and the morphological asymmetry of the BMRs can be explained by the Coriolis force acting on the Lorentz-force-driven flows.
   Furthermore, we demonstrate that the solar-like magnetic cycles can be successfully obtained if the Joy's law is explicitly given in the Babcock-Leighton $\alpha$-effect.
   In these cyclic dynamo simulation, a strong Lorentz force feedback leads to cycle modulations in the differential rotation (torsional oscillation) and meridional circulation. 
   The simulations however do not include radiative effects such as the enhanced cooling by faculae, that are required to properly model the torsional oscillations. 
   The non-axisymmetric components of the flows are found to exist as inertial modes such as the equatorial Rossby modes.
   }
  {}


   \maketitle
%

\section{Introduction}

The Sun exhibits an $11$-year cyclic magnetic activity which is sustained by the dynamo processes in the convection zone \citep[e.g.,][]{charbonneau2020}.
The Babcock-Leighton flux-transport model is one of the most promising solar dynamo models at present that can explain many observational features \citep[e.g.,][]{dikpati1999}. 
In this model, the equatorward migration of sunspot groups is attributed to the meridional flow near the base of the convection zone \citep[][]{wang1991,choudhuri1995}.
This model is supported by the recent helioseismic observations in which the meridional flow is found to be poleward at the surface and equatorward at the base \citep[][]{gizon2020}.
Another characteristic feature of this dynamo model is that the main process generating poloidal fields from toroidal  fields is  the so-called Babcock-Leighton mechanism, in which the surface poloidal fields are generated by the poleward advection and equatorial cancellation of the bipolar sunspots that are tilted with respect to east-west direction \citep{babcock1961,leighton1964}.

The tendency that the leading spot is located closer to the equator than the following one is called Joy's law \citep[][]{hale1919}.
The physical origin of the Joy's law is still under debate:
Thin flux tube simulations have demonstrated that the Joy's law can be explained by the Coriolis force acting on the buoyantly-rising flux tubes through the convection zone \citep[][]{dsilva1993,fan1994,weber2011}.
On the other hand, recent observations have shown that the active regions emerge with east-west alignment (with zero tilt) on average and the Joy's law tilts are generated by the north-south separation motions after emergence \citep[][]{schunker2020}.

Numerical investigations of the Babcock-Leighton flux-transport dynamo model have been mostly carried out in a two-dimensional (2D) kinematic mean-field framework \citep[e.g.,][]{dikpati1999,nandy2001,chatterjee2004,hazra2014,karak2016}.
In these models, the Babcock-Leighton $\alpha$-effect is modeled as the axisymmetric poloidal source term which is localized near the surface.
Although there are some non-kinematic studies where the dynamo-generated fields are allowed to give a feedback on the mean flows \citep[][]{rempel2006,ichimura2017,inceoglu2017}, the longitudinal component of the Lorentz force has been ignored because of the axisymmetry of the system. 

There are several recent studies that aim to realize the Babcock-Leighton process in a three-dimensional (3D) full-spherical domain.
\citet{yeates2013} first presented a kinematic model in which the upward velocity perturbation associated with the magnetic buoyancy is explicitly prescribed to produce the tilted bipolar magnetic regions (BMRs) at the surface.
This method has also been used in \citet{kumar2019} and \citet{whitebread2019}.
Furthermore, \citet{miesch2014} have developed a different model of the Babcock-Leighton dynamo, in which the BMRs are artificially placed at the surface in response to the toroidal field at the base under the constraint of Joy's law.
In fact, this method is regarded as a 3D realization of the so-called ``double-ring'' algorithm used in 2D mean-field models \citep[][]{durney1997,nandy2001,munoz2010}.
The same model has also been used to study the long-term cycle variability \citep{karak2017}.
However, all of these models are restricted to kinematic regime.
Therefore, it still remains unclear how the Lorentz-forces of the BMRs affect their post-emergence evolution and the resulting dynamo solution in the non-kinematic regime.

The models which include the most physics are provided by magnetohydrodynamic (MHD) convective dynamo simulations in a spherical shell \citep[e.g.,][]{brun2004,ghizaru2010,brown2010,fan2014,hotta2016,strugarek2017}.
However, they have difficulty in reproducing the large-scale mean flows as we observe when the solar parameters are used \citep[this problem is known as the convective conundrum, e.g.,][]{nelson2018}.
Moreover, they still cannot capture the full dynamics of the flux-emergence and the resulting formation of BMRs at the surface comprehensively \citep[][]{nelson2011,fan2014,chen2017}.
Therefore, it is still helpful to use mean-field\footnote{
In the solar dynamo community, the ``mean-field'' models are conventionally regarded as 2D axisymmetric models where the mean is taken over longitudes. In this paper, however, we use the term ``mean-field'' in a more general sense; the mean should be regarded as an ensemble average or a spatial average over small portions in the convection zone that satisfy the Reynolds' averaging rules. See \citet[][\S 2.1 therein]{pipin2022} for more details.}
 models in which the large-scale mean-flows are largely controlled with parameterizations of  small-scale convective angular momentum transport  \citep[the $\Lambda$-effect; see][]{kitchatinov1995} and the flux emergence is modeled via a parametrization.


In this paper, we present a new numerical framework to study the Babcock-Leighton dynamo processes of the Sun in a 3D non-kinematic regime, which takes advantage of both the mean-field approach for the solar large-scale mean flows and the 3D realization of the Babcock-Leighton process.
Therefore, our model extends both the 2D non-kinematic mean-field models \citep[e.g.,][]{rempel2006} and the 3D kinematic models \citep[e.g.,][]{miesch2016}.
Although our model is still less complete than 3D MHD convective dynamo models, we can solve the MHD dynamo equations under the constraints of the observed differential rotation and meridional circulation (that are hard to obtain in the 3D MHD convective dynamo models).
We believe that our model can potentially provide many future applications such as data assimilation and cycle prediction.

We note that a similar 3D mean-field model was recently presented by \citet[][]{pipin2022} in which the mean-field induction equation is solved together with the mean-field hydrodynamic equations.
Although the model considers the non-axisymmetric magnetic field such as newly-emerged BMRs, all the hydrodynamic variables were assumed to be axisymmetric and only the longitudinally-averaged Lorentz force was taken into account.
In our model, by contrast, we will consider the non-axisymmetric Lorentz force feedback on the non-axisymmetric flows, which we find crucial for the post-emergence evolution of the BMRs.

Recently, various kinds of inertial modes have been discovered and identified on the Sun \citep[e.g.,][]{loeptien2018,gizon2021}.
Since these inertial modes have different mode properties from those of acoustic (p) modes, they are expected to be useful as an alternative tool to probe the interior of the Sun \citep[][]{gizon2021,bekki2021a}.
However, it remains largely uncertain how these modes are affected by the dynamo-generated magnetic fields.
We believe that our 3D non-kinematic dynamo model can also be used to study the effects of magnetic fields on various inertial modes in the Sun's convection zone in the nonlinear regime.

The organization of this paper is as follows.
The numerical model is explained in detail in \S \ref{sec:model}.
In \S~\ref{sec:paramstudy}, we show how the post-emergence evolution of the BMRs are changed from the previous models.
Our initial results of the cyclic dynamo are then presented in \S\ref{sec:results}.
We close by summarizing our results and discussing the future prospects in \S\ref{sec:summary}.


\section{Model} \label{sec:model}


\subsection{Governing Equations} \label{sec:eqs}
We numerically solve a set of MHD equations in a spherical coordinate $(r,\theta,\phi)$:
\begin{eqnarray}
&& \frac{\partial \rho_{1}}{\partial t} = -\nabla\cdot (\rho_{0} \bm{v}), \label{eq:mass} \\
&& \frac{\partial \bm{v}}{\partial t} = -\bm{v}\cdot\nabla\bm{v}-\frac{\nabla p_{1}}{\rho_{0}}-\frac{\rho_{1}}{\rho_{0}}g\bm{e}_{r}+2\bm{v}\times\bm{\Omega_{0}} \nonumber \\
&& \ \ \ \ \ \ \ \ \ \ \ 				+\frac{1}{4\pi\rho_{0}} (\nabla\times\bm{B})\times\bm{B}+\frac{1}{\rho_{0}}\nabla\cdot\bm{\Pi}, \label{eq:motion} \\
&& \frac{\partial \bm{B}}{\partial t} = \nabla \times (\bm{v}\times \bm{B}+\bm{\mathcal{E}}-\eta\nabla\times\bm{B}), \label{eq:induction} \\
&& \frac{\partial s_{1}}{\partial t} = \bm{v}\cdot\nabla s_{1}+c_{\mathrm{p}}\delta\frac{v_{r}}{H_{p}}+\frac{1}{\rho_{0}T_{0}}\nabla\cdot (\rho_{0}T_{0}\kappa\nabla s_{1}) \nonumber \\ 
&& \ \ \ \ \ \ \ \ \ \ \ 	+\frac{1}{\rho_{0}T_{0}}\left[(\bm{\Pi}\cdot\nabla)\cdot\bm{v} +\frac{\eta}{4\pi}|\nabla\times\bm{B}|^{2} \right], \label{eq:entropy}
\end{eqnarray}
where $g$, $\rho_{0}$, $p_{0}$, and $H_{p}$ denote the gravitational acceleration, density, pressure, and pressure scale height of the background state which is in an adiabatically-stratified hydrostatic equilibrium.
We use the same radial profiles for the background stratification as the model presented in \citet{rempel2005} and \citet{bekki2017a}.
The quantities with subscript $1$, $\rho_{1}$ and $p_{1}$, are the perturbations with respect to the background that are assumed to be sufficiently small, i.e., $|p_{1}/p_{0}| \approx |\rho_{1}/\rho_{0}| \ll 1$, so that the equation of state is linearized
\begin{eqnarray}
&& p_{1}=p_{0}\left( \gamma\frac{\rho_{1}}{\rho_{0}} +\frac{s_{1}}{c_{v}}\right),
\end{eqnarray}
where $\gamma=5/3$ is the specific heat ratio and $s_{1}$ is entropy perturbation from the adiabatic background.
The rotation rate of the radiative core $\Omega_{0}/2\pi=431.3$ nHz is used for a system rotation rate.

The tensor $\bm{\Pi}$ represents the turbulent Reynolds stress associated with small-scale (subgrid-scale) convective motions that are not explicitly resolved in our model.
This in principle contains the effects of turbulent diffusion and turbulent momentum transport \citep[$\Lambda$-effect, see][]{kitchatinov1995}.
Therefore, the Reynolds stress is expressed as,
\begin{eqnarray}
&&\Pi_{ik}=\rho_{0} \nu_{\mathrm{vis}} \left( S_{ik}-\frac{2}{3}\delta_{ik}\nabla\cdot\bm{v}  + \Lambda_{ik}\Omega_{0} \right),
\end{eqnarray}
where $S_{ik}$ and $\delta_{ik}$ denote the velocity deformation tensor and the Kronecker-delta unit tensor.
The detailed expression of $S_{ik}$ in a spherical coordinate can be found in \citet{fan2014}.

In our model, turbulent viscous, thermal, and magnetic diffusivities are all assumed to be isotropic.
We use the same radial profiles for the viscous ($\nu_{\mathrm{vis}}$), thermal ($\kappa$), and magnetic ($\eta$) diffusivities as of \citet{rempel2006}.
The magnetic diffusivity $\eta$ is $10^{12}$ cm$^{2}$ s$^{-1}$ at the top boundary and is on the order of $10^{11}$ cm$^{2}$ s$^{-1}$ in the bulk of the convection zone \citep[see Fig.~2 in][]{rempel2006}.
Although this diffusivity value is smaller than the estimate by the local mixing-length model \citep[][]{munoz2011}, it is still about one order magnitude larger than the typical diffusivity value used in the kinematic flux-transport models in the advection-dominated regime \citep[e.g.,][]{dikpati1999,guerrero2007}.

In order to break the Taylor-Proudman's constraint of the differential rotation via the thermal wind balance, 
a negative (positive) latitudinal entropy gradient in the northern (southern) hemisphere is required.
Although there are several proposed mechanisms to generate this latitudinal entropy gradients \citep[e.g.,][]{kitchatinov1995,masada2011,hotta2018}, in this paper, we adopt the idea proposed by \citet{rempel2005} that the latitudinal entropy variation is generated by the radial meridional flows when the base of the convection zone is weakly subadiabatic.
To this end, we give the superadiabaticity $\delta=\nabla-\nabla_{\mathrm{ad}}$, with $\nabla=d\ln{T}/d\ln{p}$, as
\begin{eqnarray}
&& \delta(r,\theta) =\mathcal{T}_{-}(r;r_{\mathrm{sub}},d_{\mathrm{sub}}) \ \delta_{\mathrm{sub}}(\theta), \\
&& \delta_{\mathrm{sub}}(\theta) = \delta_{\mathrm{pl}}+(\delta_{\mathrm{eq}}-\delta_{\mathrm{pl}}) \sin^{2}{\theta}, \\
&& r_{\mathrm{sub}}(\theta)=r_{\mathrm{pl}}+(r_{\mathrm{eq}}-r_{\mathrm{pl}})\sin^{2}{\theta},
\end{eqnarray}
where $\mathcal{T}$ denotes a transition function defined by
\begin{eqnarray}
&& \mathcal{T}_{\pm}(x;x_{0},d)=\frac{1}{2}\left[1\pm\tanh{\left( \frac{x-x_{0}}{d}\right)} \right].
\end{eqnarray}
We set the superadiabaticity at the poles and at the equator as $\delta_{\mathrm{pl}}=-1.5\times 10^{-5}$ and $\delta_{\mathrm{eq}}=-2\times 10^{-5}$, respectively, at the base of the convection zone.
The depths where the stratification changes from subadiabatic to adiabatic are given as $r_{\mathrm{pl}}=0.725R_{\odot}$ and $r_{\mathrm{eq}}=0.735R_{\odot}$.
The weakly subadiabatic layer near the base is thought to be an outcome of a non-local energy transport of strongly magnetized convection \citep{skaley1991,brandenburg2016} and has been reported in some recent numerical simulations \citep{kapyla2017,hotta2017,bekki2017b}.
The subadiabaticity is slightly enhanced in the equatorial area owing to the latitudinal variation of the Coriolis force acting on low-entropy downdrafts \citep{karak2018}.


The dimensionless tensor $\Lambda_{ik}$ specifies the amplitude and direction of the turbulent momentum transport.
In this model, we only consider the turbulent angular momentum transport.
Therefore, we parameterize $\Lambda_{r\phi}(=\Lambda_{\phi r})$ and $\Lambda_{\theta\phi}(=\Lambda_{\phi\theta})$ similarly to the model presented in \citet{rempel2005},
\begin{eqnarray}
&& \Lambda_{r\phi}=+\Lambda_{0}\tilde{f}_{l}(r,\theta) \cos{(\theta+\lambda)} \left[1+\zeta_{r}(r,\theta,\phi) \right],  \label{eq:lam_r} \\
&& \Lambda_{\theta\phi}=-\Lambda_{0}\tilde{f}_{l}(r,\theta) \sin{(\theta+\lambda)} \left[1+\zeta_{\theta}(r,\theta,\phi) \right]. \label{eq:lam_q}
\end{eqnarray}
The overall amplitude of the $\Lambda$-effect is given by $\Lambda_{0}=0.85$.
The inclination is set to $\lambda=+(-)15^{\circ}$ in the northern (southern) hemisphere.
Thus, the associated angular momentum flux becomes largely equatorward and weakly away from the rotational axis.
The spatial distribution of the $\Lambda$-effect is specified as
\begin{eqnarray}
&& \tilde{f}_{l}(r,\theta)=\frac{f_{l}(r,\theta)}{\mathrm{max}|f_{l}(r,\theta)|}, \\
&& f_{l}(r,\theta)=\sin^{2}{\theta}\cos{\theta} \tanh{\left( \frac{r_{\mathrm{max}}-r}{d_{l}}\right)}.
\end{eqnarray}
where $d_{l}=0.025R_{\odot}$.
With this parameterization, the profiles of differential rotation and meridional circulation become similar to observations \citep{howe2009,gizon2020}

The quantities $\zeta_{r}$ and $\zeta_{\theta}$ denote random fluctuations due to the unresolved turbulent convection.
In our model, the random fields $\zeta_{r}$ and $\zeta_{\theta}$ are separately constructed by simply superposing multiple gaussians as,
\begin{eqnarray}
 \zeta(r,\theta,\phi)=\sum_{i=1}^{N} c_{i} \exp{\left[ -\left( \frac{r-r_{i}}{\delta r}\right)^{2}-\left( \frac{\theta-\theta_{i}}{\delta \theta}\right)^{2}
	-\left( \frac{\phi-\phi_{i}}{\delta \phi}\right)^{2}  \right]},
\end{eqnarray}
where the locations of gaussian peaks $(r_{i},\theta_{i},\phi_{i})$ are randomly chosen and their amplitudes $c_{i}$ are also randomly determined within the range $-2<c_{i}<2$.
The spatial scale of each gaussian is set as $(\delta r,\delta\theta,\delta\phi)=(0.03R_{\odot},5^{\circ},5^{\circ})$.
In our reference calculation, we set the number of gaussians $N=30$.
We generate the random field $\bm{\zeta}$ at every time step and therefore it is uncorrelated in time.
Note that the non-axisymmetric flows can be partially driven by these random fluctuations of the $\Lambda$-effect.


In the induction equation (\ref{eq:induction}), we add an electro-motive-force $\bm{\mathcal{E}}$ to model the Babcock-Leighton $\alpha$-effect, by which the poloidal field is generated near the surface from the toroidal field near the base of the convection zone.
Note that this term is only switched on when the cyclic dynamo is simulated in \S~\ref{sec:results}.
A detail formulation of $\bm{\mathcal{E}}$ will be given in \S~\ref{sec:BLalpha}.

\subsection{Numerical scheme} \label{sec:scheme}

We numerically solve the Eqs.~(\ref{eq:mass})--(\ref{eq:entropy}) using the $4$th-order centered-difference method for space and $4$-step Runge-Kutta scheme for time integration \citep{voegler2005}.
To avoid the severe CFL constraint for time step, we use the reduced speed of sound technique \citep[][]{rempel2005} so that the background sound speed is artificially reduced by a factor of $\xi=200$, which still ensures that flows remain sufficiently subsonic \citep[][]{hotta2014}.
Moreover, we use the hyperbolic divergence cleaning method (9-wave method) for minimizing the numerical error resulting from the divergence of magnetic field \citep{dedner2002}.

The numerical domain is a full-spherical shell extending from $r_{\mathrm{min}}=0.65R_{\odot}$ up to $r_{\mathrm{max}}=0.985R_{\odot}$. 
The base of the convection zone is located at $r_{\mathrm{bc}}=0.71R_{\odot}$.
In order to avoid the singularities in a spherical coordinate at the poles, we use the Yin-Yang grid \citep{kageyama2004}.
For more details about the implementation of the Yin-Yang grid, refer \citet{bekki2021b}.
The grid resolution is $72(N_{r})\times72(N_{\theta})\times216(N_{\phi})\times2$(Yin and Yang grids).
The code is parallelized using message passing interface (MPI).
At both radial boundaries, impenetrable and stress-free boundary condition is used for velocity.
The magnetic field is assumed to be radial at the top and horizontal at the bottom.

\begin{figure}[]
\begin{center}
\includegraphics[width=0.9\linewidth]{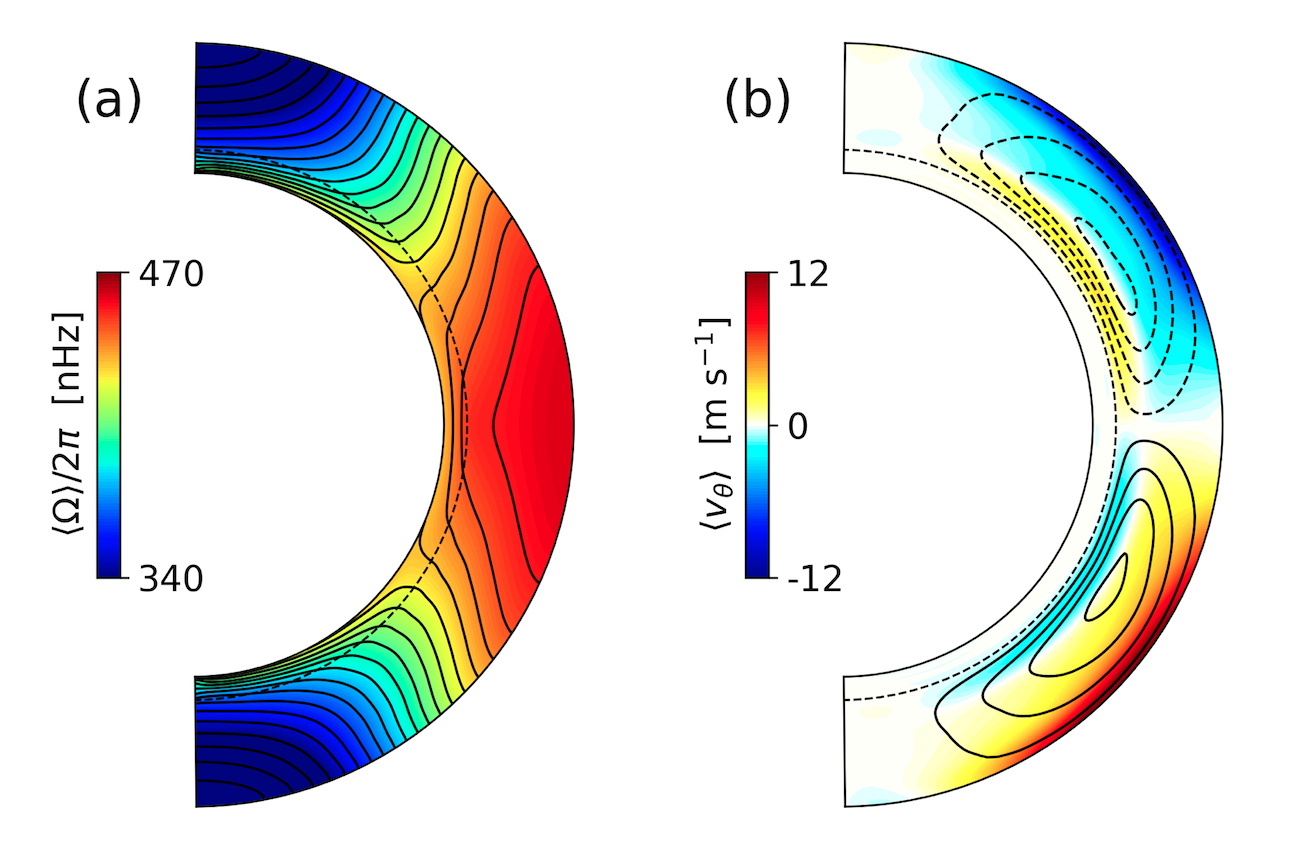}
\caption{
Results of the mean flows from the hydrodynamic calculation.
(a) Differential rotation $\langle \Omega\rangle =\Omega_{0}+\langle v_{\phi} \rangle /(r\sin{\theta})$.
The black dotted curve shows the location of the base of the convection zone at $r=0.71R_{\odot}$, below which the stratification is subadiabatic.
(b) Meridional circulation $\langle v_{\theta} \rangle$.
The black solid and dashed curves show contours of the streamfunction $\Psi$ defined by $\rho_{0} \bm{v_{m}}=\nabla\times(\Psi \bm{e}_{\phi})$ where $\bm{v}_{m}=\langle v_{r} \rangle \bm{e}_{r} +\langle v_{\theta} \rangle \bm{e}_{\theta}$.
The meridional circulation is counter-clockwise (clockwise) in the northern (southern) hemisphere, i.e., the flow is poleward (equatorward) at the surface (base).
}
\label{fig:icHD}
\end{center}
\end{figure}

We first carry out a hydrodynamic simulation until the large-scale mean flows become quasi-stationary.
Figure~\ref{fig:icHD} shows profiles of the differential rotation $\langle \Omega \rangle = \Omega_{0}+\langle v_{\phi} \rangle/(r\sin{\theta})$ and meridional circulation $\bm{v}_{m}=\langle v_{r}\rangle \bm{e}_{r}+\langle v_{\theta}\rangle \bm{e}_{\theta}$ obtained from our hydrodynamic simulation.
Here, $\langle \ \rangle$ denotes the longitudinal average.
We then add magnetic fields to carry out MHD calculations.


\section{Post-emergence evolution of BMRs} \label{sec:paramstudy}

\begin{table}[]
 \begin{center} 
\caption{Model parameters of post-emergence BMRs simulations.}
\small
\vspace{-0.3\baselineskip}
\begin{tabular}{cccccccccc} 
\toprule
\toprule
  \renewcommand{\arraystretch}{1.8}
Case  & & $\Delta r_{\mathrm{bmr}}/R_{\odot}$ & $\Delta \phi_{\mathrm{bmr}}$ [deg] & Shape & \\
 \midrule
1 &...& $0.04$ & $15$ & horizontal    \\
2 &...& $0.08$ & $10$ & round    \\
3 &...& $0.12$ &  $5$ & vertical   \\
\bottomrule
\end{tabular}
\label{table:1}
\end{center}
\vspace{-1.0\baselineskip}
\tablefoot{
The subsurface shape of a BMR changes from a horizontally-elongated half-ellipse in Case 1 to a vertically-elongated half-ellipse in Case 3.
In Case 2, the subsurface field structure is close to round-shaped.
} 
\end{table}

In this section, we carry out a set of numerical simulations to study how the post-emergence evolution of a BMR is dependent on how it is injected into the simulation.
To this end, we solve the MHD equations (\ref{eq:mass})--(\ref{eq:entropy}) starting from different initial magnetic field configurations for a newly-emerged single BMR.
For simplicity, we only consider the short-term evolution of a BMR and do not discuss the long-term buildup of the polar fields and the resulting dynamo cycles.
Hence, we set $\bm{\mathcal{E}}=0$.

\begin{figure*}[]
\centering
\includegraphics[width=0.96\linewidth]{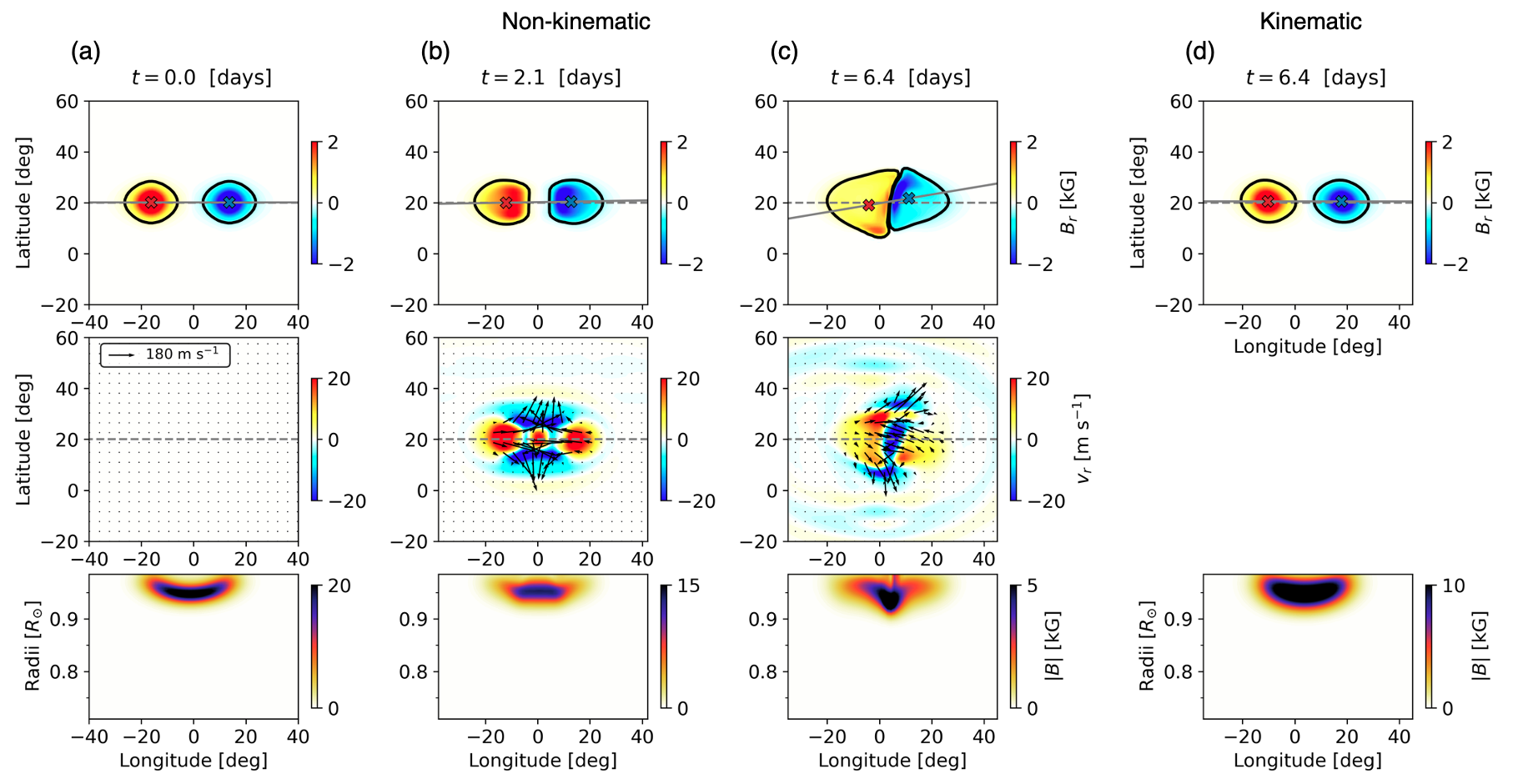}
\caption{
Temporal evolution of a BMRs from Case 1 at selected temporal points (a) $t=0$ days, (b) $t=2.1$ days, and (c) $t=6.4$ days.
The kinematic simulation with the same initial condition is shown in panel (d) at $t=6.4$ days.
Top panels show the radial field $B_{r}$ at the surface $r=0.985R_{\odot}$.
Thick black solid curves shows the contour at $|B_{r}|=0.3$ kG.
The blue and red cross marks represent the locations of the flux-weighted center for the leading and following spots, and the grey straight lines are drawn to connect these two cross marks.
Middle panels show the radial flows $v_{r}$ near the surface $r=0.98R_{\odot}$ (color contour) and the horizontal velocities ($v_{\theta},v_{\phi}$) at the surface (vector arrows).
Bottom panels show cross sections of the magnetic field strength $|B|$ at the fixed latitude of $20^{\circ}$ which is denoted by black dashed lines in the top and middle panels.
}
\label{fig:results-case1}
\end{figure*}
\begin{figure*}[]
\centering
\includegraphics[width=0.85\linewidth]{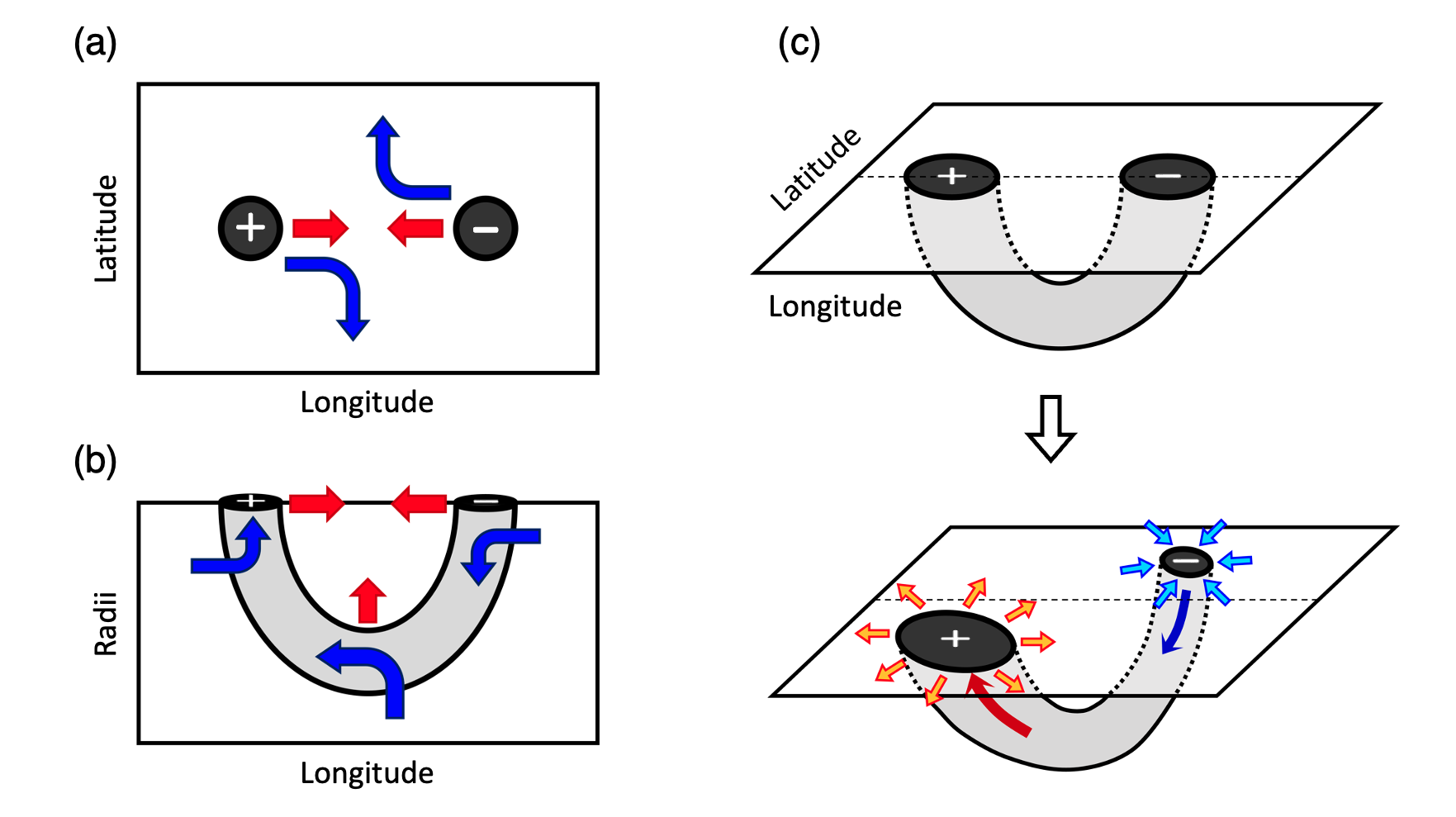}
\caption{
Schematic illustrations explaining the generation of anti-Joy's law tilt and the morphological asymmetry of the BMRs' field strengths.
(a) Cross section at the surface seen from the top. The red and blue arrows show the directions of the Lorentz force and the Coriolis force, respectively.
(b) Cross section at the fixed latitude seen from the equator to the north pole.
(c) Three-dimensional view of the evolution of a BMRs.
}
\label{fig:schematic}
\end{figure*}

\begin{figure*}[]
\centering
\includegraphics[width=0.96\linewidth]{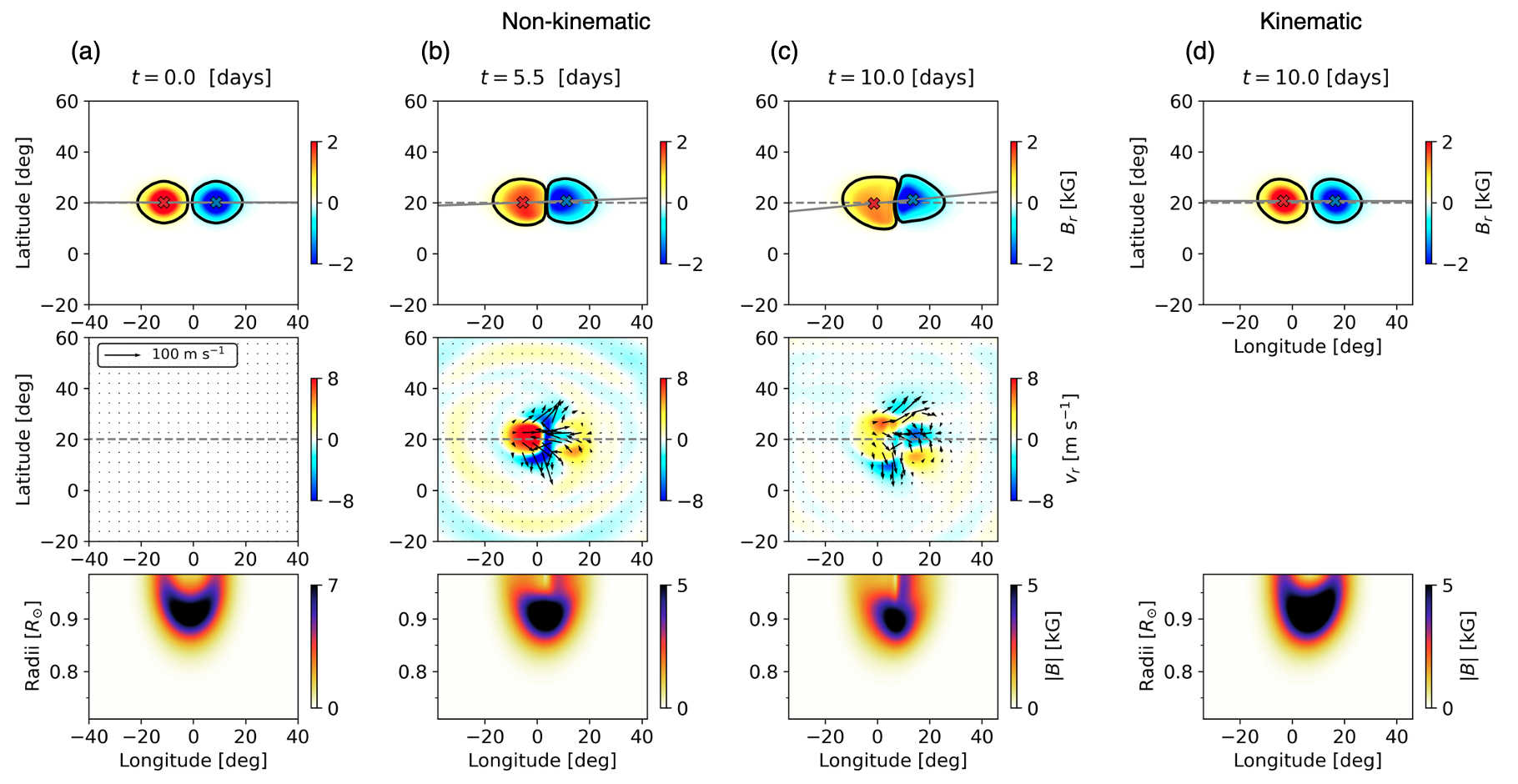}
\caption{
Same as Fig.~\ref{fig:results-case1} except from the simulation Case 2.
}
\label{fig:results-case2}
\end{figure*}
\begin{figure*}[]
\centering
\includegraphics[width=0.96\linewidth]{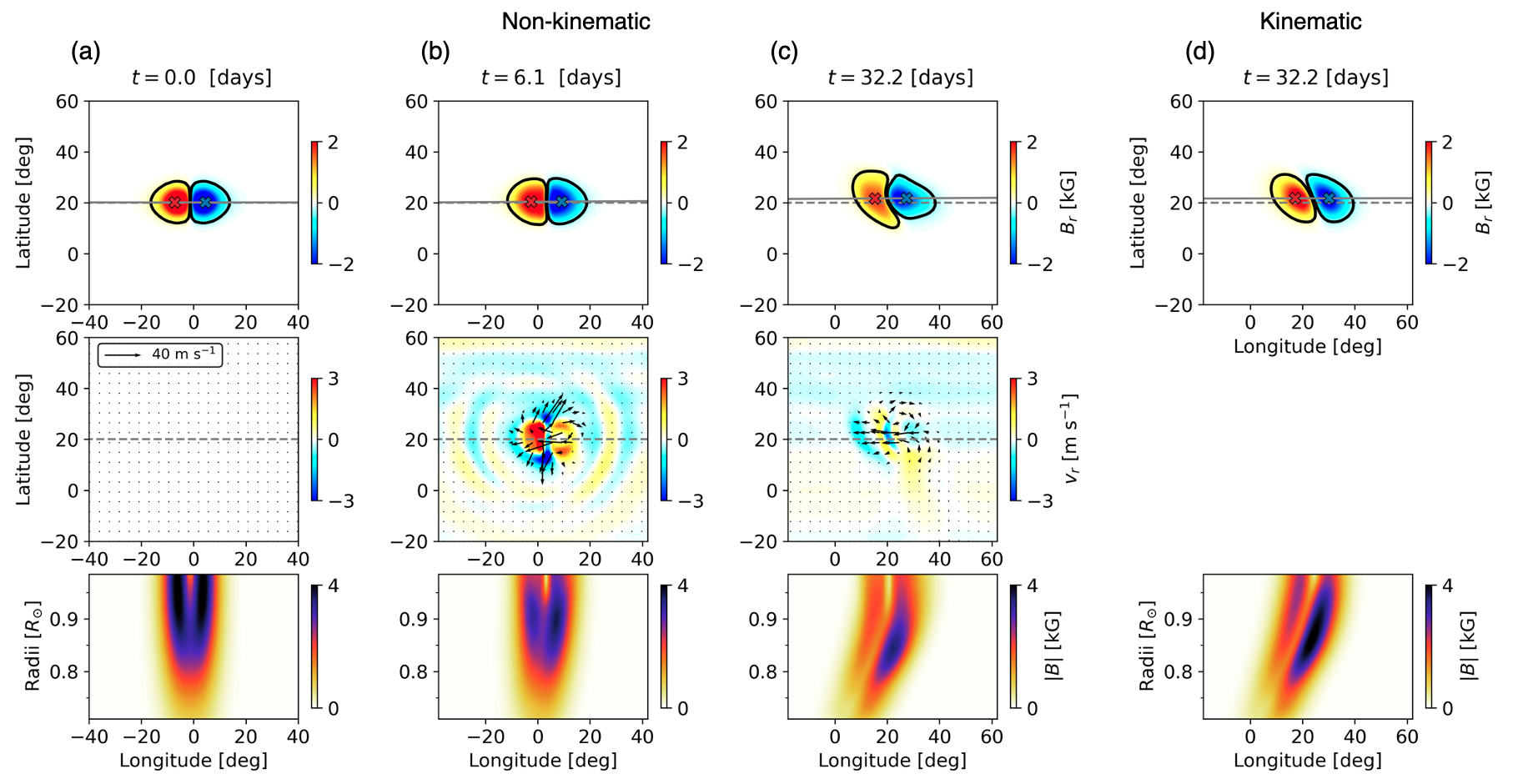}
\caption{
Same as Fig.~\ref{fig:results-case1} except from the simulation Case 3.
}
\label{fig:results-case3}
\end{figure*}

\subsection{Initial condition of magnetic fields}

For simplicity, we simulate the evolution of a BMR with zero initial tilt, i.e., the leading and following polarity spots are perfectly east-west aligned.
The initial magnetic field is given as
\begin{eqnarray}
    && \bm{B} \propto \nabla \times (A_{\mathrm{ic}} \bm{e}_{\theta}), \label{eq:Bic1}
\end{eqnarray}
where the vector potential $A_{\mathrm{ic}}$ is given by
\begin{eqnarray}
    A_{\mathrm{ic}}(r,\theta,\phi)=\left\{ 1-\tanh{\left( \frac{l_{\mathrm{bmr}}(r,\phi)-1}{0.5} \right)}\right\} 
    \exp{\left[ -\left( \frac{\theta-\theta^{*}}{\Delta\theta_{\mathrm{bmr}}}\right)^{2}\right]}, \label{eq:Bic2}
\end{eqnarray}
with 
\begin{eqnarray}
    && l_{\mathrm{bmr}}(r,\phi)=\sqrt{\left( \frac{r-r_{\mathrm{max}}}{\Delta r_{\mathrm{bmr}}}\right)^{2}+\left( \frac{\phi-\phi^{*}}{\Delta\phi_{\mathrm{bmr}}}\right)^{2}}. \label{eq:Bic3}
\end{eqnarray}
Here we set the colatitude $\theta^{*}=70^{\circ}$ and $\phi^{*}=0^{\circ}$, and therefore a BMR is located at the latitude of $20^{\circ}$ in the northern hemisphere at the central meridian.
The parameters $\Delta r_{\mathrm{bmr}}$, $\Delta \theta_{\mathrm{bmr}}$, and $\Delta \phi_{\mathrm{bmr}}$ specify the radial, latitudinal, and longitudinal extent of the BMR.
We fix $\Delta \theta_{\mathrm{bmr}}=5^{\circ}$ but vary both $\Delta r_{\mathrm{bmr}}$ and $\Delta \phi_{\mathrm{bmr}}$ as free model parameters to change the subsurface shape of the BMR, as summarized in Table.~\ref{table:1}.
In Case 1, $\Delta r_{\mathrm{bmr}}$ is relatively small and $\Delta \phi_{\mathrm{bmr}}$ is large, indicating that the subsurface structure of the BMR is very shallow in radius but stretched in longitude.
In Case 3, on the other hand, the subsurface field morphology is changed to a vertically-elongated half ellipse.
Case 2 is an intermediate case between Case 1 and 3.
In all cases, the amplitude of the initial field is determined such that the maximum radial field at the top boundary is $4$ kG.

\subsection{Temporal evolution} \label{sec:temporalevolution}

Let us first take a look at Case 1 where the results are most drastically changed from the previous studies.
Figure~\ref{fig:results-case1} shows the evolution of the radial field, flows at the surface, and the subsurface field of the BMR over the first several days after the emergence from Case 1.
As soon as the BMR is inserted, there emerge strong upflows at the surface because the mass is expelled from the flux tube due to the pressure imbalance.
At the same time, the strong Lorentz force of the BMR drives  strong longitudinal converging flows towards the polarity inversion line and latitudinal diverging flows along the polarity inversion line.
These are clearly seen in Fig.~\ref{fig:results-case1}b middle panel.
Consequently, the two spots (that are initially separated in longitude) are quickly pulled together and stretched in latitude, as shown in Fig.~\ref{fig:results-case1}c top panel.
For comparison, we show the same snapshot from the corresponding kinematic calculation in as Fig.~\ref{fig:results-case1}d.
The result reveals that the temporal evolution of the BMR is substantially changed in non-kinematic regime where the Lorentz force and the Coriolis force are taken into account.
This type of evolution is not observed on the Sun.

\subsection{Tilt angle} \label{tilt}

In order to measure the tilt angle of the BMR, we compute the flux-weighted center locations of the leading and following polarity regions ($\theta_{L}$, $\phi_{L}$) and ($\theta_{F}$, $\phi_{F}$), respectively.
The tilt angle $\gamma$ is then defined as
\begin{eqnarray}
\gamma &=&\tan^{-1}{\left( \frac{\sin{\theta_{L}}-\sin{\theta_{F}}}{\cos{\theta_{L}}\sin{\theta_{L}}-\cos{\theta_{F}}\sin{\theta_{F}}} \right)} \nonumber \\
& \approx &  \tan^{-1}{\left[ \frac{\theta_{L}-\theta_{F}}{(\phi_{L}-\phi_{F}) \cos{\theta^{*}}} \right]}.
\end{eqnarray}
Since we consider the BMR located in the northern hemisphere, the Joy's law is satisfied when $\gamma >0$ by definition.
It is seen from Fig.~\ref{fig:results-case1}c (top) that the leading polarity spot is located at slightly higher latitude than the following polarity spot on average, indicating that the associated tilt angle is negative ($\gamma \approx -10^{\circ}<0$ at $t=6.2$ days).
This is against the Joy's law.
This is because the Coriolis force acts on the longitudinal converging flows (towards the polarity inversion line), as schematically illustrated in Fig.~\ref{fig:schematic}a.
It should be emphasized that this generation of anti-Joy's law tilt is essentially 3D non-kinematic effect, and thus, cannot be captured neither by the 2D non-kinematic models (which ignores the longitudinal dependence) nor the 3D kinematic models (which ignores the Lorentz force feedback).

Needless to say, the generation of negative tilts is contrary to the solar observations.
One possible way of reconciling these simulations and observations is that, the Joy's law tilts are generated during the rise of the toroidal flux tubes \citep[e.g.,][]{dsilva1993} and thus are already embedded apriori in the emerging BMRs, which overcomes the tendency to generate the anti-Joy's law tilts.
In \S~\ref{sec:results}, we will demonstrate that this is possible.
Another possibility is that the BMRs emerge with nearly zero tilt but acquire the positive tilts after the emergence by yet-unknown physical process. 
For example, \citet{martinbelda2016} proposed that the net positive tilt can be generated from the initial zero-tilt state due to the coupled effects of differential rotation and active region inflows.
This effect is not considered in this study since our code does not include the effect of radiative cooling in the BMRs, which geostrophically drives the active region inflows \citep[e.g.,][]{spruit2003}.

\subsection{Morphological asymmetry} \label{asymmetry}

We also find a significant asymmetry between the leading and following spots:
The leading polarity region tends to retain its compact shape and its strong field strength, whereas the following spot tends to gradually expand and the field becomes substantially weaker as time passes.
To qualitatively assess this field strength asymmetry, we measure the maximum field strengths in the leading and following polarity regions $B_{L}$ and $B_{F}$.
In our simulation Case 1, the leading spot has about twice stronger field than the following spot, $|B_{L}/B_{F}|\approx 2$ at $t=6.2$ days.
This morphological asymmetry of the BMRs has been a well-known feature observed on the Sun \citep[e.g.,][]{bray1979,fisher2000}, and often explained by the differential stretching of the rising $\Omega$-loop due to the Coriolis force \citep[e.g.,][]{fan1993}. 
See \citet[][]{fan2021} for a more comprehensive review on the observational and theoretical studies on the morphological asymmetry.

Here, we provide a different but related explanation for this asymmetry.
As illustrated in Fig.~\ref{fig:schematic}b, the two spots are attracted with each other by the Lorentz force, and the Coriolis force acting on these longitudinal converging flows drives an downflow (upflow) inside the leading (following) polarity regions of the flux tube.
This can be confirmed by the contour map of the radial motion in Fig.~\ref{fig:results-case1}c (middle).
Owing to the mass conservation, these downflow and upflow are accompanied by the horizontal converging and diverging motions, respectively.
Therefore, the leading polarity region becomes compressed and the field gets stronger, whereas the following one become broader and the field gets weaker.
This is schematically illustrated in Fig.~\ref{fig:schematic}c.

\begin{figure}[]
\centering
\includegraphics[width=0.85\linewidth]{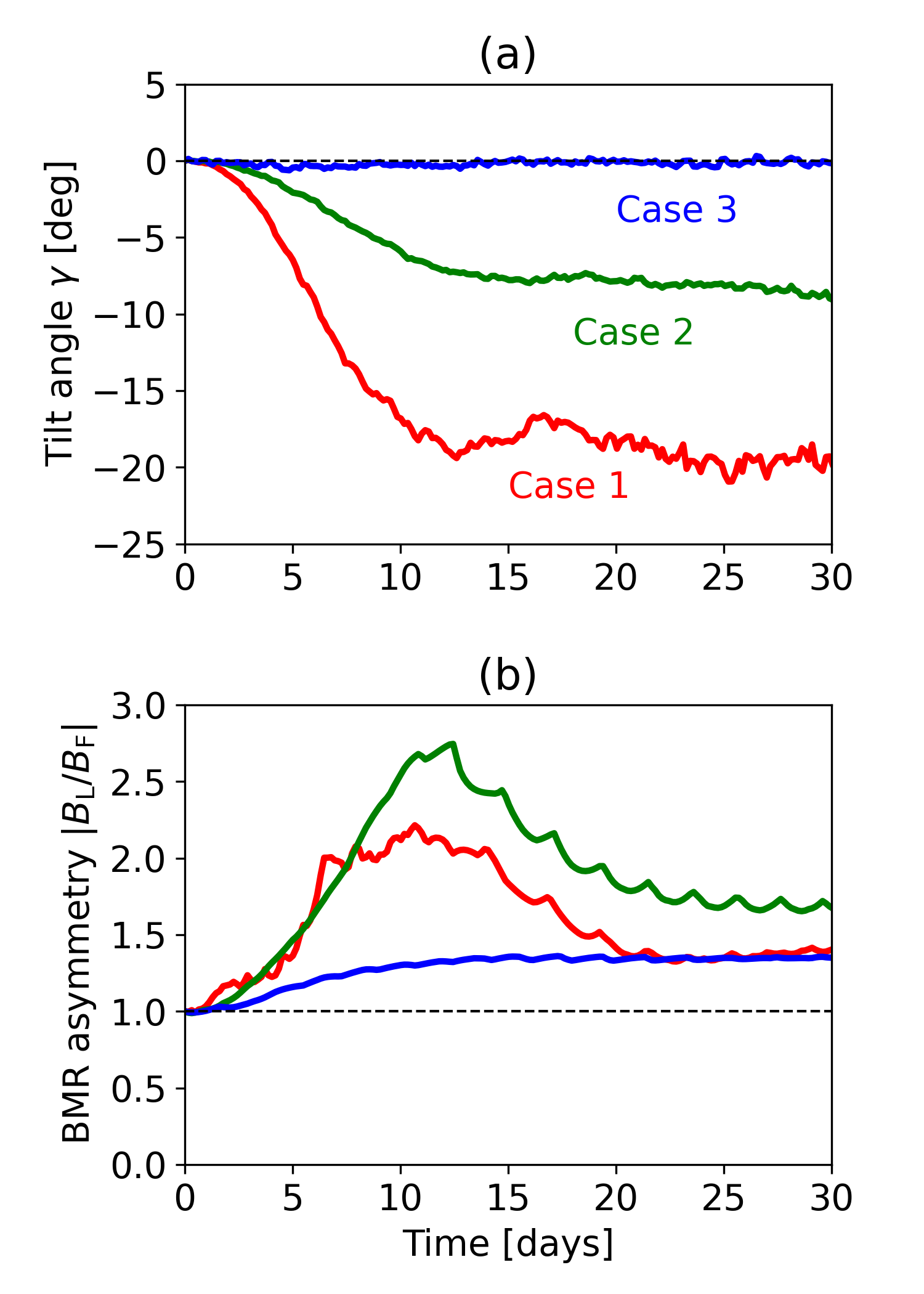}
\caption{
Temporal evolution of (a) the tilt angle $\gamma$ and (b) the ratio of the maximum field strengths between the leading and following magnetic regions $|B_{\mathrm{L}}/B_{\mathrm{F}}|$.
The Red, green, and blue curves correspond to the Cases 1, 2, and 3.
}
\label{fig:tile_time}
\end{figure}

\subsection{Dependence on the subsurface structure of a BMR}

The results from the Case 2 and Case 3 are shown in Fig.~\ref{fig:results-case2} and Fig.~\ref{fig:results-case3}, respectively.
We can clearly see that the temporal evolution of a BMR is sensitively dependent upon the initial field structure of the BMR.
From Case 1 (where the BMR is localized in a shallow surface and the two spots are largely separated in longitude) to Case 3 (where the BMR extends deeper in the convection zone and the longitudinal separation of the spots is small), 
the dominant component of the magnetic tension force is changed from horizontal (longitudinally converging at the surface) to vertical (upward at the bottom apex of the flux tube).
Since the longitudinal component of the Lorentz force at the surface is dominated by the magnetic tension force, the surface driving of the longitudinal converging flows decreases from Case 1 to case 3.
This is explained in more detail in Appendix~\ref{app:A}.
Consequently, the generation of a negative tilt is significantly suppressed from Case 1 to Case 3, as shown in Fig.~\ref{fig:tile_time}a.

On the other hand, the field strength asymmetry between the leading and following spots can be found in all cases as shown in Fig.~\ref{fig:tile_time}b.
This is because, in the Case 3, the Lorentz-force-driven rising motion of the bottom apex of the flux tube drives the counter-rotating flow inside the flux tube, which can enhance the converging (diverging) motion in the leading (following) polarity regions at the surface.
Therefore, regardless of the subsurface shape of the BMR, the observed morphological asymmetry can be reproduced.
In fact, the asymmetry is most significant in Case 2 where the Coriolis force can act both on the longitudinally converging motions at the surface and on the radially upward motion of the deep flux tube.
This is schematically illustrated in Figs.~\ref{fig:schematic}b and c.

Although the simulations reported in this section are based on a very simplified model of the half-torus-shaped BMR with zero initial tilt, 
we find that its temporal evolution is extremely sensitive to its subsurface structure in the 3D non-kinematic regime.
In particular, this study warns that the shallow BMRs model (which is conventionally used in many 2D non-kinematic models and 3D kinematic models) will lead to drastically different dynamo results when all these realistic 3D non-kinematic effects are included, especially due to the tendency to produce the anti-Joy's law tilts.


\section{Cyclic dynamo with Babcock-Leighton $\alpha$-effect} \label{sec:results}

In the previous section, we see that the newly-emerged BMRs with shallow subsurface root have a general tendency to produce the anti-Joy's law tilts in the 3D non-kinematic regime.
In this section, we demonstrate that, despite this trend, the cyclic dynamo solution can be obtained if the Joy's law tilt is explicitly imposed for the emerging BMRs.

\subsection{Babcock-Leighton $\alpha$-effect source term} \label{sec:BLalpha}

Now, we switch on the electro-motive-force $\bm{\mathcal{E}}$ in the Eq.~(\ref{eq:induction}) that represents the source of the Babcock-Leighton $\alpha$-effect, by which the surface poloidal field is produced as a result of the north-south tilt of the BMRs \citep{babcock1961,leighton1964}.
In our model, the emergence of BMRs at the surface is assumed to occur in response to the dynamo-generated toroidal field deep inside the convection zone, i.e.,
the tilted BMRs are instantaneously generated when the toroidal field near the base exceeds a threshold field strength.
Our approach differs from the method presented in \citet{yeates2013}, \citet{kumar2019}, and \citet{pipin2022} where the upward velocity associated with the magnetic buoyancy of the toroidal flux is prescribed.
Rather, our method is similar to that used in \citet{miesch2014} and \citet{miesch2016} where the BMRs are explicitly spotted at the surface.

We take the following steps to construct $\bm{\mathcal{E}}$.
First, the toroidal field near the base of the convection zone is computed at every time step,
\begin{eqnarray}
&& \bar{B}_{\mathrm{tor}}(\theta,\phi)=\frac{1}{r_{b}-r_{a}}\int_{r_{a}}^{r_{b}} B_{\phi}(r,\theta,\phi) dr,
\end{eqnarray}
where the average is taken over a narrow radial range near the base of the convection zone from $r_{a}=0.71R_{\odot}$ to $r_{b}=0.735R_{\odot}$.
Then, we determine the location of the flux emergence in a spherical surface $(\theta^{*},\phi^{*})$.
In order to suppress the emergence at high latitudes as suggested by observations, we apply a latitudinal mask to $ \bar{B}_{\mathrm{tor}}(\theta,\phi)$ such that
\begin{eqnarray}
&& \bar{B}^{*}_{\mathrm{tor}}(\theta,\phi)=\mathcal{T}_{+}(\theta;\pi/2-\theta_{\mathrm{em}},\Delta\theta_{\mathrm{tran}})\times  \nonumber \\
&&	 \ \ \ \ \ \ \ \ \ \ \ \ \ \ \ \ \ \ \ \  	 \mathcal{T}_{-}(\theta;\pi/2+\theta_{\mathrm{em}},\Delta\theta_{\mathrm{tran}})  \bar{B}_{\mathrm{tor}}(\theta,\phi),
\end{eqnarray}
where $\theta_{\mathrm{em}}=17.5^{\circ}$ and $\Delta \theta_{\mathrm{tran}}=8.5^{\circ}$.
We impose a necessary condition for the BMRs emergence to occur, that $|\bar{B}^{*}_{\mathrm{tor}}(\theta,\phi)|$ exceeds a threshold field strength $B_{\mathrm{crit}}=750$ G.
When the above condition is satisfied on multiple points, the location of emergence $(\theta^{*},\phi^{*})$ is randomly chosen between points satisfying the criterion.

Eventually, $\bm{\mathcal{E}}$ is expressed as follows being proportional to $\bar{B}^{*}_{\mathrm{tor}}(\theta^{*},\phi^{*})$,
\begin{eqnarray}
&& \left(
    \begin{array}{c}
      \mathcal{E}_{r} \\
      \mathcal{E}_{\theta} \\
      \mathcal{E}_{\phi}
    \end{array}
  \right)=a_{0} \tilde{f}^{*}_{\alpha} (r,\theta,\phi) \left(
    \begin{array}{c}
      0 \\
      -\cos{\psi^{*}} \\
      \sin{\psi^{*}}
    \end{array}
  \right)  \bar{B}_{\mathrm{tor}}^{*}(\theta^{*},\phi^{*}), \label{eq:emf}
  \end{eqnarray}
where $\tilde{f}_{\alpha}$ represents the spatial distribution of BMRs,
\begin{eqnarray}
 \tilde{f}^{*}_{\alpha}(r,\theta,\phi) = \exp{\left[ -\left( \frac{r-r_{\mathrm{max}}}{\Delta r_{\mathrm{bmr}}}\right)^2
            -\left( \frac{\theta-\theta^{*}}{\Delta\theta_{\mathrm{bmr}}}\right)^2 -\left( \frac{\phi-\phi^{*}}{\Delta\phi_{\mathrm{bmr}}}\right)^2  \right]}.
\end{eqnarray}
Here, $\Delta r_{\mathrm{bmr}}$, $\Delta \theta_{\mathrm{bmr}}$, and $\Delta \phi_{\mathrm{bmr}}$ denote the radial, latitudinal, and longitudinal size of the BMRs.
In order to demonstrate that the cyclic dynamo is possible even with the presence of strong non-kinematic effects (discussed in \S~\ref{sec:paramstudy}), we set $\Delta r_{\mathrm{bmr}}=0.04R_{\odot}$.
Therefore, the BMRs are confined in the shallow surface layer, which is necessary to avoid the poleward dynamo wave propagation (see Appendix~\ref{app:B2}).
In the reference calculation, we set $\Delta\theta_{\mathrm{bmr}}=\Delta\phi_{\mathrm{bmr}}=6^{\circ}$, which is consistent with observations suggesting the typical size of BMRs of $r_{\mathrm{bmr}}\approx 5-100$ Mm \citep[e.g.,][]{solanki2003} that implies $\Delta\theta_{\mathrm{bmr}}=\Delta\phi_{\mathrm{bmr}}\approx 2r_{\mathrm{bmr}}/R_{\odot}\approx 0.4-8^{\circ}$.
The overall amplitude of the Babcock-Leighton $\alpha$-effect is set to $a_{0}=100$ km s$^{-1}$.
This value, in combination with the typical toroidal field strength near the base $ \bar{B}_{\mathrm{tor}}^{*}\approx 5-20$ kG \citep[][]{dikpati1999}, leads to the total magnetic flux of BMRs of $10^{22}-10^{23}$ Mx, which is consistent with observations \citep[][]{schrijver1994}.

\begin{figure}
\begin{center}
\includegraphics[width=0.95\linewidth]{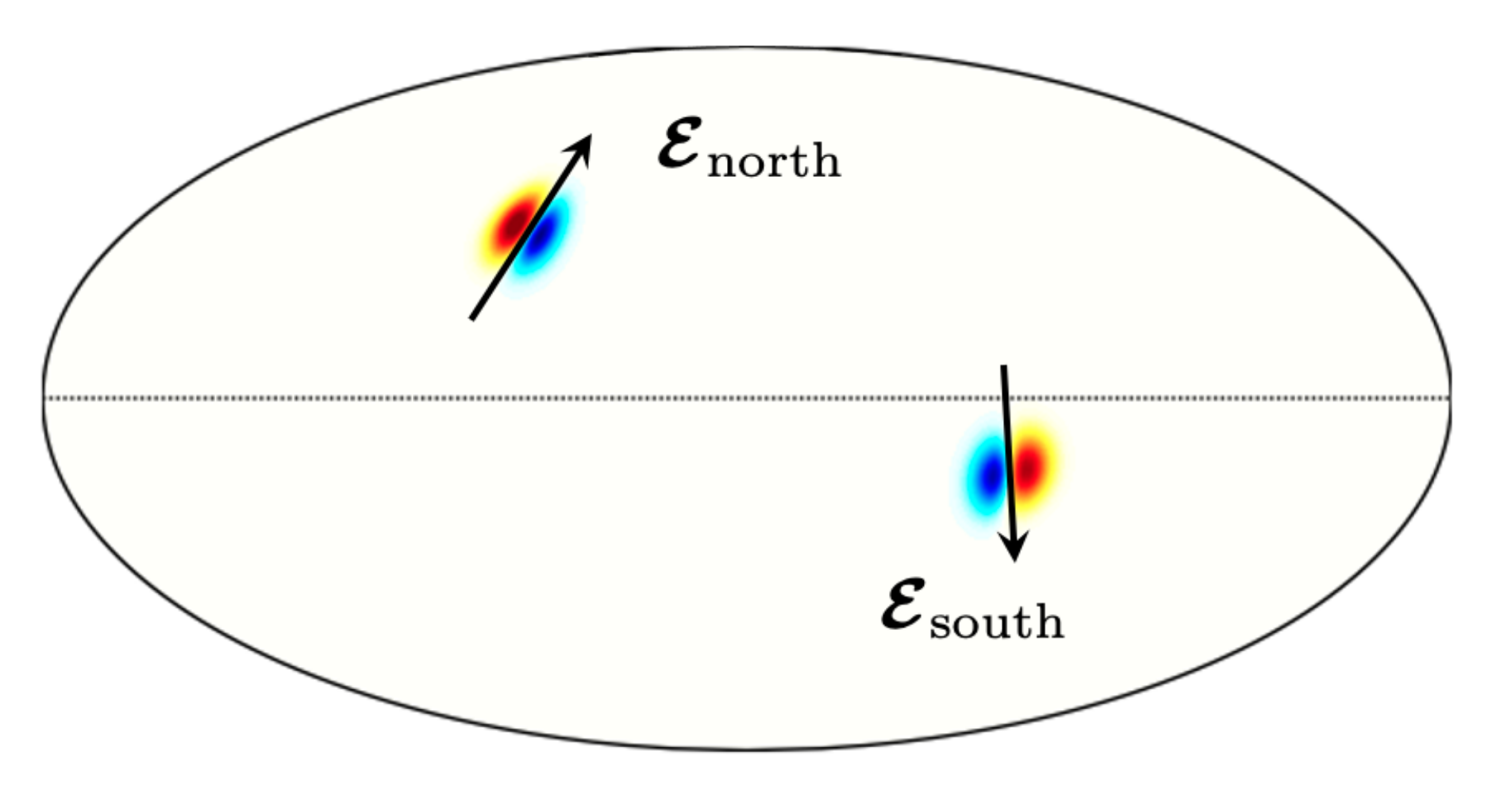}
\caption{Example structure of BMRs per each hemisphere produced from our Babcock-Leighton $\alpha$-effect model.
Radial field at the solar surface is shown where red (blue) points represent positive (negative) $B_{r}$.
The Solid black arrows denote the direction of the electro-motive-force $\mathcal{E}$ defined by the Eq.(\ref{eq:emf}) with appropriate Joy's law.
Positive (negative) toroidal field in the northern (southern) hemisphere is implicitly assumed near the base of the convection zone.}
\label{fig:BLemf}
\end{center}
\end{figure}

The north-south tilt of the BMRs ($\psi^{*}$) obeys the Joy's law such that
\begin{eqnarray}
&& \psi^{*}=35^{\circ}\cos{\theta^{*}}+\psi'_{\mathrm{f}},
\end{eqnarray}
where $\psi'_{\mathrm{f}}$ denotes the random fluctuation of the tilt angle \citep{hale1919,howard1991,stenflo2012,wang2015}.
For simplicity, we assume that the probability distribution of $\psi'_{\mathrm{f}}$ is roughly given by the following Gaussian distribution,
\begin{eqnarray}
&& P_{\mathrm{f}}(\psi'_{\mathrm{f}})=\frac{1}{\sigma_{\mathrm{f}}\sqrt{2\pi}}\exp{\left[ -\psi_{\mathrm{f}}^{\prime 2}/(2\sigma_{\mathrm{f}}^{2}) \right]},
\end{eqnarray}
with $\sigma_{\mathrm{f}}=15^{\circ}$.
Unlike the kinematic model of \citet{karak2017}, a quenching term is not necessary in our model because the saturation of the dynamo occurs self-consistently owing to the Lorentz-force feedback \citep{rempel2006,ichimura2017}.
Figure~\ref{fig:BLemf} shows examples of $\bm{\mathcal{E}}$ and the resulting tilted BMRs produced by our Babcock-Leighton $\alpha$-effect source where we assume sufficiently strong positive (negative) toroidal field near the base of the convection zone.

\begin{figure}
\begin{center}
\includegraphics[width=0.95\linewidth]{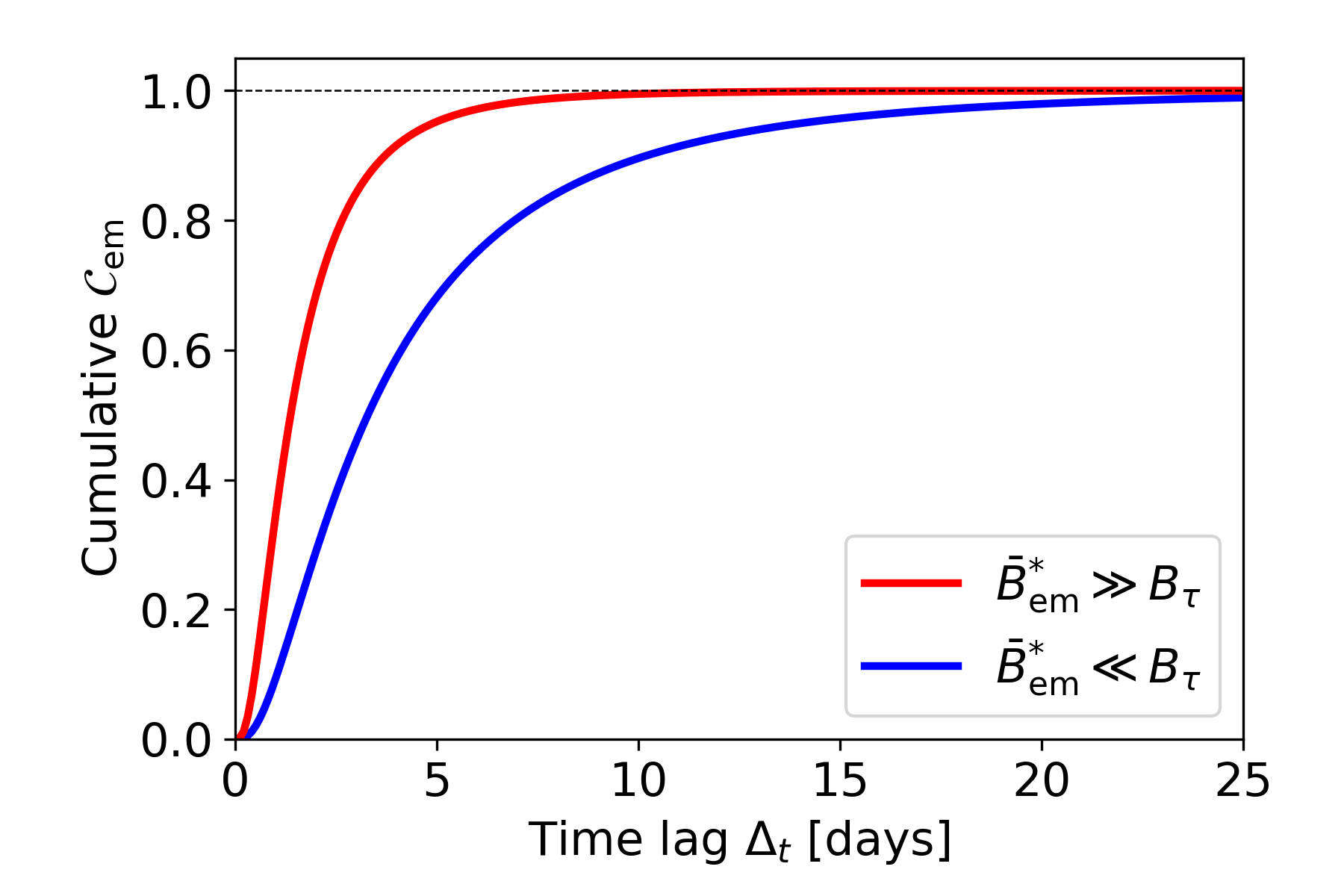}
\caption{
Cumulative log-normal distribution of the emergence events $\mathcal{C}_{\mathrm{em}}(\Delta_{t})$ used in our model during the activity maxima (red) and activity minima (blue).
}
\label{fig:cumulative}
\end{center}
\end{figure}

In order to prevent overlapping emergence events on the same location in a very short time span, we introduce the following time delay algorithm as presented in \citet{miesch2014,miesch2016,karak2017}:
We use a cumulative log-normal distribution function of the emergence events defined as
\begin{eqnarray}
&& \mathcal{C}_{\mathrm{em}}(\Delta_{t})=\int_{t=t_{s}}^{t_{s}+\Delta_{t}} P_{\mathrm{em}}(\Delta_{t}) \ dt, \\
&& P_{\mathrm{em}}(\Delta_{t})=\frac{1}{\sigma_{t}\Delta_{t}\sqrt{2\pi} }\exp{\left[ -\frac{(\ln{\Delta_{t}}-\mu_{t})^{2}}{2\sigma_{t}^{2}}\right]},
\end{eqnarray}
where $\Delta_{t}=t-t_{s}$ is the time lag since the last emergence event at $t_{s}$.
A flux emergence event is allowed only when the cumulative $\mathcal{C}_{\mathrm{em}}$ exceeds a number $\in [0,1]$ randomly chosen at every time step.
The standard deviation $\sigma_{t}$ and the mean $\mu_{t}$ are specified by $\tau_{p}$ and $\tau_{s}$ as follows.
\begin{eqnarray}
&& \sigma_{t}^{2}=\frac{2}{3}\ln{\left( \frac{\tau_{s}}{\tau_{p}}\right)}, \ \ \ 
\mu_{t}=\sigma_{t}^{2}+\ln{\tau_{p}}, \\
&& \tau_{p}=\tau_{p,0}+ \Delta \tau_{p} \ e^{-(\bar{B}^{*}_{\mathrm{em}}/B_{\tau})^{2}}, \\
&& \tau_{s}=\tau_{s,0}+ \Delta \tau_{s} \ e^{-(\bar{B}^{*}_{\mathrm{em}}/B_{\tau})^{2}}.
\end{eqnarray}
Here, we set $\tau_{p,0}=0.8$ days, $\tau_{s,0}=1.9$ days, $\Delta\tau_{p}=0.75$ days, $\Delta\tau_{s}=3.0$ days.
The quantity $\bar{B}^{*}_{\mathrm{em}}$ represents the horizontally-averaged $\bar{B}^{*}_{\mathrm{tor}}$ and $B_{\tau}=1.5$ kG denotes the threshold value of the toroidal field strength for characterizing phase in the activity cycle.
Figure~\ref{fig:cumulative} shows the two examples of the cumulative $\mathcal{C}_{\mathrm{em}}(\Delta_{t})$ each corresponding to the solar activity minima and maxima.
Therefore, the flux emergence becomes more frequent during the activity maxima ($\bar{B}^{*}_{\mathrm{em}} > B_{\tau}$) and less frequent when during the activity minima ($\bar{B}^{*}_{\mathrm{em}} < B_{\tau}$).

We must note that our Babcock-Leighton $\alpha$-effect model is strongly spatially-localized and temporally intermittent.
This is clearly different from the conventional 2D models with spatially-distributed and temporally-continuous source term \citep[][]{choudhuri1995,dikpati1999,rempel2006}.
Taking into account the tilt angle inclination of $\bm{\mathcal{E}}$, the localization in longitudes, and the emergence frequency of the BMRs, we can estimate the corresponding $\alpha_{0}$ value within the 2D mean-field framework as
\begin{eqnarray}
\alpha_{0} &\approx& \sin{\psi}\left(\frac{\Delta \phi_{\mathrm{bmr}}}{2\pi} \right) \left( \frac{\Delta t_{\mathrm{CFL}}}{\Delta_{t}} \right) a_{0} \nonumber \\
&\approx& 1.2 \ \mathrm{m}\ \mathrm{s}^{-1},
\end{eqnarray}
where we use the typical values $\psi=17.5^{\circ}$, $\Delta_{t}=5$ days, and $\Delta t_{\mathrm{CFL}}=17$ min.
This $\alpha_{0}$ value is consistent with the previous 2D mean-field models.

\begin{figure*}[t]
\begin{center}
\includegraphics[width=\linewidth]{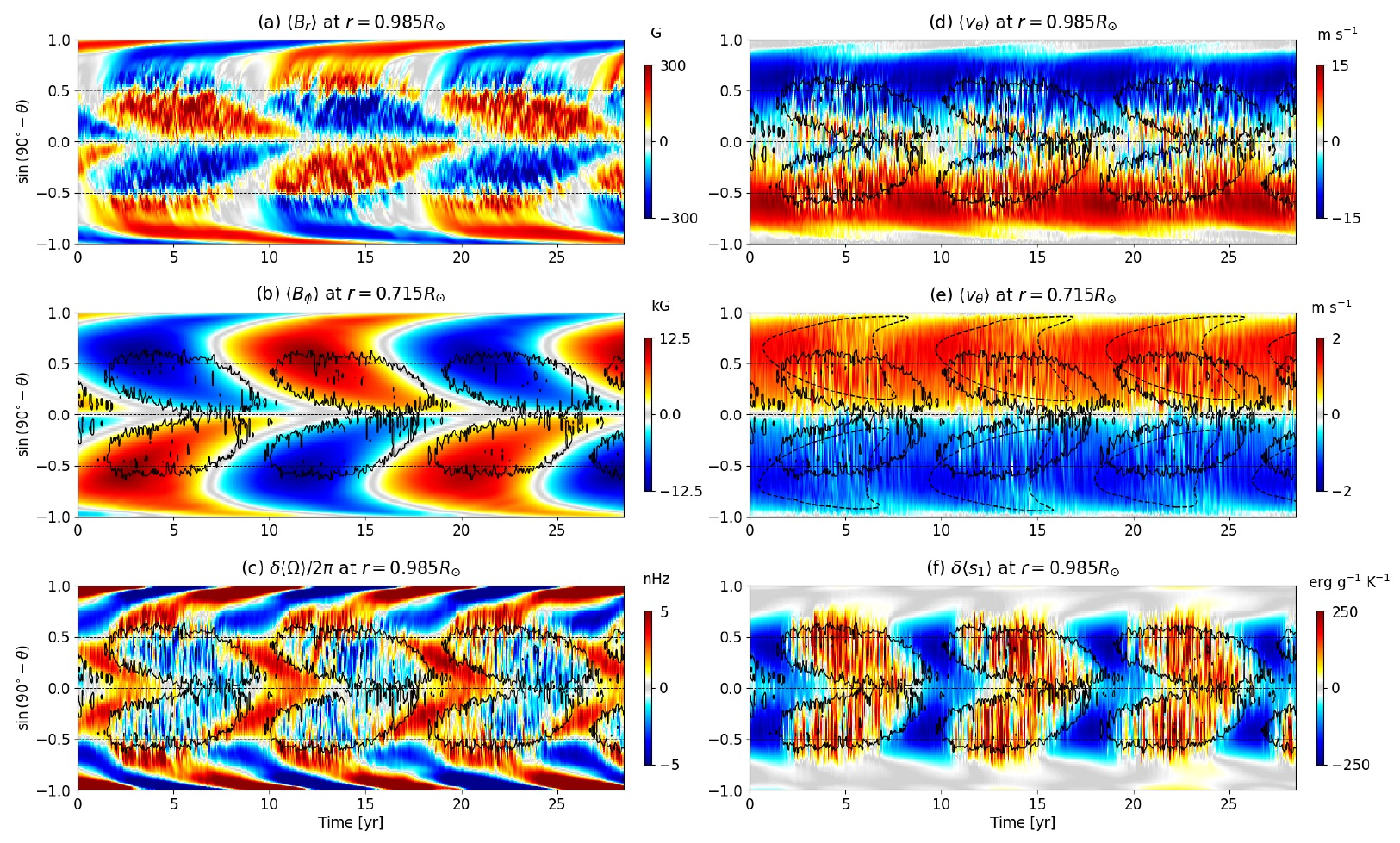}
\caption{Temporal evolution of the longitudinally-averaged magnetic fields and horizontal velocities.
(a) Azimuthal mean of the radial field $\langle {B}_{r} \rangle$ at the surface $r=0.985R_{\odot}$ where the bar denotes the azimuthal mean.
(b) Azimuthal mean of the longitudinal field $\langle {B}_{\phi} \rangle$ near the base of the convection zone $r=0.715R_{\odot}$.
Black solid lines are the contours of the emerged BMRs at each time.
(c) Torsional oscillation pattern $\delta \langle \Omega \rangle = \langle \Omega \rangle -\langle\Omega \rangle_{t}$ at the surface where $\langle \rangle_{t}$ denotes the azimuthal and temporal average.
(d) Azimuthal mean of the latitudinal velocity $\langle {v}_{\theta} \rangle$ at the surface.
Red (blue) in the northern hemisphere represents the equatorward (poleward) flow.
(e) The same as (d) but near the base of the convection zone.
Black dashed lines denote the contours of the toroidal field at the base (8.5 kG).
(f) Entropy perturbation $\delta \langle s_{1} \rangle = \langle s_{1} \rangle-\langle s_{1}\rangle_{t}$ at the surface.
}
\label{fig:bfly}
\end{center}
\end{figure*}

\begin{figure}[]
\begin{center}
\includegraphics[width=0.93\linewidth]{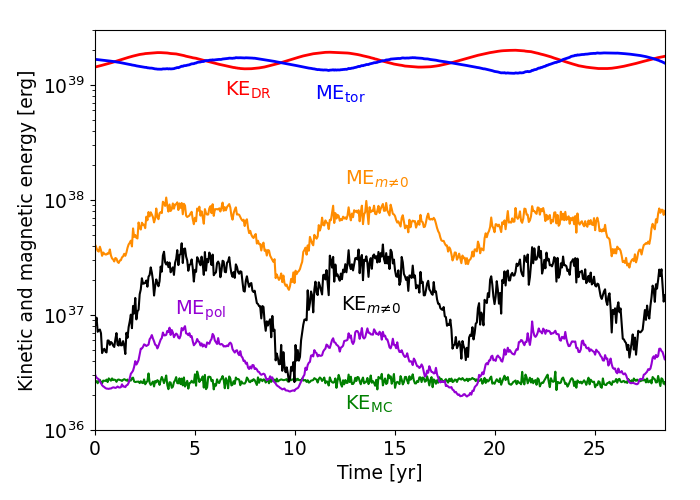}
\caption{
Temporal evolution of the volume-integrated kinetic and magnetic energies.
The red, green, and black lines denote the kinetic energies of the differential rotation $\mathrm{KE}_{\mathrm{DR}}$, the meridional circulation $\mathrm{KE}_{\mathrm{MC}}$, and the non-axisymmetric flows $\mathrm{KE}_{m \ne 0}$.
The blue, purple, and orange lines denote the magnetic energies of the mean toroidal field $\mathrm{KE}_{\mathrm{tor}}$, the mean poloidal field $\mathrm{ME}_{\mathrm{pol}}$, and the non-axisymmetric fields $\mathrm{ME}_{m \ne 0}$.
}
\label{fig:Etime}
\end{center}
\end{figure}

\subsection{Initial condition of magnetic fields} 

We add an axisymmetric dipolar field into the fully-developed hydrodynamic calculation shown in Fig.~\ref{fig:icHD}.
The simulation is evolved until initial transients disappear and the dynamo cycles with quasi-steady amplitudes are obtained.
In the following subsections, we analyze the last three cycles of our simulation.

\subsection{Dynamo cycles and the Lorentz force feedback}

Figures~\ref{fig:bfly}a and b show the tim-latitude plots of the longitudinally-averaged radial field $\langle {B}_{r} \rangle$ at the surface and the toroidal field $\langle {B}_{\phi} \rangle$ near the base of the convection zone, represented in terms of the well-known magnetic butterfly diagram. 
We can clearly see the cyclic polarity reversals that occur roughly at every 9 years, which is slightly shorter than the solar cycle yet comparable.
In each cycle, there is an equatorward migration of sunspot groups (BMRs) and the build-up of the polar field by poleward advection of the magnetic fluxes associated with the trailing sunspots.
These are owing to the single-cell meridional circulation achieved in our model, which has an amplitude of about $15$ m s$^{-1}$ at the surface and $2$ m s$^{-1}$ near the base of the convection zone.
The black solid lines in Fig.~\ref{fig:bfly}b denote the range of the emergence latitudes of BMRs at each time (the so-called active region belt).
The phase of the equatorward advection of the toroidal field at the base corresponds to that of the emergence of the BMRs at the surface.

In our non-kinematic model, the dynamo-generated fields have strong impacts on flows via the Lorentz force feedback.
Figure~\ref{fig:bfly}c shows the time-latitude plot of the fluctuation of the differential rotation $\delta \langle \Omega \rangle = \langle \Omega\rangle -\langle \Omega\rangle_{t}$ where $\langle \ \rangle_{t}$ denotes the longitudinal and temporal average.
This is commonly known as torsional oscillations \citep[][]{howard1980}.
We clearly find both poleward and equatorward propagating oscillation patterns with the typical amplitude of about $5$ nHz at the surface.
Figures~\ref{fig:bfly}d and e show the time-latitude plots of the latitudinal velocity $\langle v_{\theta}\rangle$ at the top and bottom of the convection zone, respectively.
Although the poleward flow at the surface and the equatorward flow near the base are strongly suppressed and disturbed during the activity maxima, the feedback is not large enough to switch off the advective transport of the magnetic fields \citep[][]{ichimura2017}.

Figure~\ref{fig:bfly}f shows the time-latitude plot of the entropy perturbation $\delta \langle s_{1} \rangle = \langle s_{1} \rangle -\langle s_{1}\rangle_{t}$ at the surface with typical variation amplitude of about $250$ erg g$^{-1}$ K$^{-1}$ which corresponds to the temperature fluctuation of about $1.4$ K.
The positive entropy fluctuation can be seen along with the active region belt, implying that the surface is heated whenever the BMRs emerge due to the strong magnetic diffusion in our model.
Note, however, that this is not likely in the real Sun: 
The surface is expected to be cooled by enhanced radiation in the BMRs, leading to lower temperature due to the radiative loss in the active region belt \citep[][]{spruit2003}.
The theories suggest that, this radiative loss at the surface can produce the low-latitude branches of the torsional oscillation by inducing the geostrophical flows around the BMRs (thermal forcing) \citep[][]{rempel2006,gizon2008}.
This effect is not included in our model.
It should be emphasized that, in our simulation reported here, the artificial heating in the active region belt may be responsible for the low-latitude torsional oscillation branches due to the thermal forcing with the opposite sign.

Figure~\ref{fig:Etime} shows the volume-integrated kinetic and magnetic energies of various components.
Their definitions are 
\begin{eqnarray}
&& \mathrm{KE}_{\mathrm{DR}}=\int_{V} \frac{\rho_{0}}{2} \langle v_{\phi} \rangle^{2} \ dV, \\
&& \mathrm{KE}_{\mathrm{MC}}=\int_{V} \frac{\rho_{0}}{2} (\langle v_{r} \rangle^{2}+\langle v_{\theta} \rangle^{2}) \ dV, \\
&& \mathrm{KE}_{m \ne 0}=\int_{V} \frac{\rho_{0}}{2} (\bm{v}-\langle \bm{v} \rangle)^{2} \ dV, \\
&& \mathrm{ME}_{\mathrm{tor}}=\int_{V} \frac{1}{8\pi} \langle B_{\phi} \rangle^{2} \ dV, \\
&& \mathrm{ME}_{\mathrm{pol}}=\int_{V} \frac{1}{8\pi} (\langle B_{r} \rangle^{2}+\langle B_{\theta} \rangle^{2}) \ dV, \\
&& \mathrm{ME}_{m \ne 0}=\int_{V} \frac{1}{8\pi} (\bm{B}-\langle \bm{B} \rangle)^{2} \ dV, 
\end{eqnarray}
where the integrals are taken over the whole volume of the convection zone.
The two largest energy reservoirs in our simulation are the differential rotation kinetic energy $\mathrm{KE}_{\mathrm{DR}}$ and the toroidal field magnetic energy $\mathrm{ME}_{\mathrm{tor}}$.
When the toroidal field is amplified by the $\Omega$-effect, $\mathrm{KE}_{\mathrm{DR}}$ is converted to $\mathrm{ME}_{\mathrm{tor}}$.
The toroidal field eventually becomes superequipartition ($\mathrm{ME}_{\mathrm{tor}}>\mathrm{KE}_{\mathrm{DR}}$) with respect to the differential rotation on average.
In our simulation, this Lorentz force feedback on differential rotation leads to a dynamo saturation.
The non-axisymmetric magnetic energy $\mathrm{ME}_{m \ne 0}$ is greater than the mean poloidal field energy $\mathrm{ME}_{\mathrm{pol}}$ because the BMRs are strongly non-axisymmetric.
The non-axisymmetric fields drive strong non-axisymmetric flows, whose kinetic energy $\mathrm{KE}_{m \ne 0}$ is also greater than $\mathrm{ME}_{\mathrm{pol}}$.
This suggests that the non-axisymmetric components of the magnetic fields and the flows are important for the convection zone dynamics.

\begin{figure*}
\begin{center}
\includegraphics[width=0.995\linewidth]{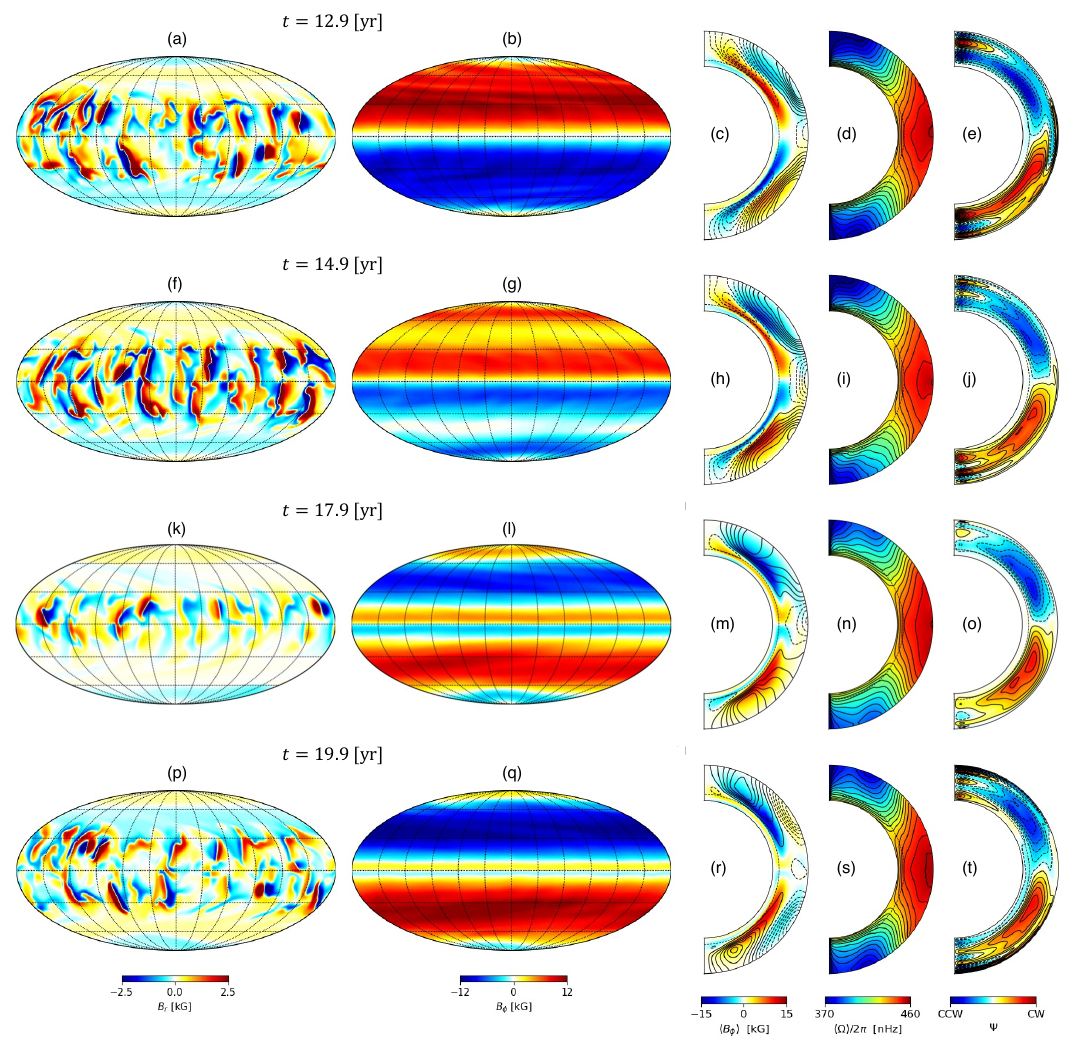}
\caption{Time evolution of magnetic field and velocity.
Shown are the snapshots at $t=12.9$ yr (from (a) to (e)), $t=14.9$ yr (from (f) to (j)), $t=17.9$ yr (from (k) to (o)), $t=19.9$ yr (from (p) to (t)) in Fig.~\ref{fig:bfly}.
The mollweide projections on the 1st and 2nd columns show the radial field $B_{r}$ at the surface $r=0.985R_{\odot}$ and longitudinal field $B_{\phi}$ near the base of the convection zone $r=0.715R_{\odot}$, respectively.
The meridional plot in the 3rd column represents the azimuthally-averaged toroidal field (color scales) and poloidal field (contours).
The meridional plots in the 4th and 5th columns represent the azimuthally-averaged differential rotation and streamfunction of the meridional circulation, respectively.
An animation of this figure is available online.
}
\label{fig:evol_vB}
\end{center}
\end{figure*}

\begin{figure*}
\begin{center}
\includegraphics[width=0.99\linewidth]{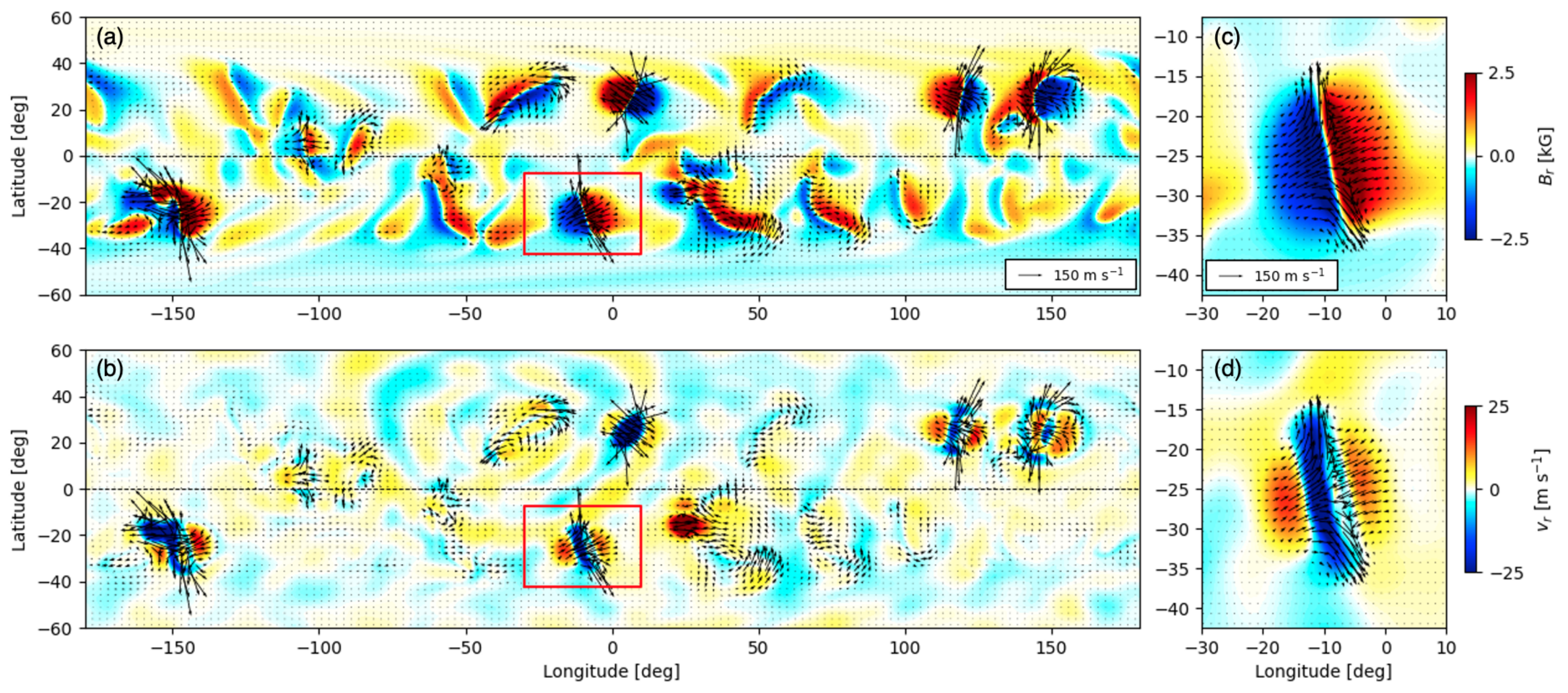}
\caption{
Snapshots of the radial field $B_{r}$ at the surface (top panels) and the radial velocity $v_{r}$ [m s$^{-1}$] near the surface (bottom panels) at $t=10.3$ yr in Fig.~\ref{fig:bfly}.
The black arrows represent the horizontal flow ($v_{\theta},v_{\phi}$) at the surface.
Panels (c) and (d) are the zoom-in of the panels (a) and (b), focusing on the single BMRs denoted by red thick solid lines.
}
\label{fig:sfvec}
\end{center}
\end{figure*}

\begin{figure}
\begin{center}
\includegraphics[width=0.99\linewidth]{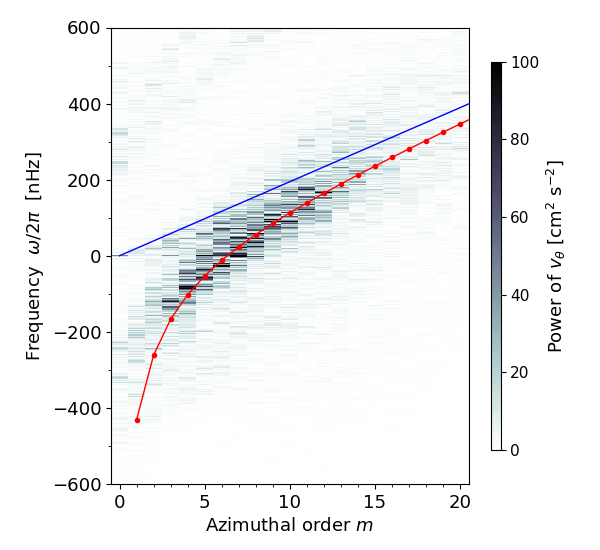}
\caption{
Equatorial power spectrum of latitudinal velocity $v_{\theta}$ near the surface $r=0.95R_{\odot}$.
The spectra are computed in a frame rotating at $\Omega_{\mathrm{0}}/2\pi=431.3$ nHz.
The blue solid lines represent the differential rotation rate at the surface $\omega=m (\Omega_{\mathrm{sf}}-\Omega_{0} )$ where $\Omega_{\mathrm{sf}}=\langle \Omega(0.95R_{\odot},\pi/2) \rangle$.
The red points denote the theoretical dispersion relation of the sectoral Rossby modes, $\omega=-2\Omega_{\mathrm{sf}}/(m+1)+m ( \Omega_{\mathrm{sf}}-\Omega_{0})$.
}
\label{fig:rossby}
\end{center}
\end{figure}

Figure~\ref{fig:evol_vB} shows snapshots of the magnetic fields and mean flows over the course of a magnetic cycle.
The two leftmost panels show the mollweide projections of the radial magnetic field $B_{r}$ at the surface and the toroidal field $B_{\phi}$ at the bottom convection zone.
As prescribed in our Babcock-Leighton source term, BMRs emerge at low latitudes obeying the Hale's and Joy's laws.
Therefore, the radial magnetic field at the surface is substantially non-axisymmetric.
On the other hand, toroidal field near the base of the convection zone is found to be almost axisymmetric.
Meridional plots on the 3rd, 4th, and 5th columns of Fig.~\ref{fig:evol_vB} show the azimuthally-mean profiles of the poloidal and toroidal magnetic fields, differential rotation, and meridional circulation, respectively.
When longitudinally averaged, the dynamo solution shows a qualitatively similar time evolution pattern as the previous 2D mean-field models \citep[e.g.,][]{rempel2006,ichimura2017}, although our model has a stronger torsional oscillation and the meridional circulation modulations which presumably depends on the radial structure we have assumed for flux emergence.

\subsection{Non-axisymmetric flows}

In our model, non-axisymmetric flows are driven largely by the non-axisymmetric Lorentz forces and only partially by the random fluctuations in the $\Lambda$-effect.
Figure~\ref{fig:sfvec} shows a snapshot of the radial field at the surface (top rows) and the radial velocity near the surface (bottom rows).
Black arrows represent the horizontal flow motions at the surface.
Strong horizontal flows exist only in the vicinity of the BMRs:
When a BMR emerge at the surface which happens instantaneously in our model, horizontal converging flows are driven towards the polarity inversion line with the typical amplitudes of about $100$ m s$^{-1}$.
This is owing to a strong magnetic tension force of the BMRs that pulls the two spots together.
This strong converging flow further drives both horizontal outflows and radial downflows along the polarity inversion line, as shown in Fig.~\ref{fig:sfvec}c and d.
Due to these strong horizontal flows at the surface, a newly-emerged BMR that initially consists of two round-shaped sunspots is quickly deformed into an elongated shape along the polarity inversion line, as seen in Fig.~\ref{fig:sfvec}a. 
This temporal evolution is similar to the simulation reported in \S~\ref{sec:temporalevolution} with horizontally-elongated BMRs (Case 1).
We find that this elongated feature of the BMRs, which is not observed on the Sun \citep[][]{driel2015}, can be greatly suppressed if the Babcock-Leighton $\alpha$-effect is much weaker (Appendix \ref{app:B1}) or if BMRs have much deeper radial extent (Appendix \ref{app:B2}).
It should also be noted that we do not include the radiative cooling associated with the active regions in our simulations. 
Thus, our model currently lacks the physics required to properly produce the observed inflows associated with active regions \citep{gizon2001,spruit2003}.

Other interesting non-axisymmetric flow features are low-frequency inertial modes of oscillation, in particular, the equatorial Rossby modes that have recently been detected on the Sun \citep[e.g.,][]{loeptien2018}.
Figure~\ref{fig:rossby} shows the equatorial power spectrum of latitudinal velocity $v_{\theta}$ near the surface from our non-kinematic dynamo simulation similarly to \citet[][]{bekki2021b}.
Note that all the spectra are computed in a frame rotating at $\Omega_{\mathrm{ref}}/2\pi=431.3$ nHz.
We can clearly see the existence of the equatorial Rossby modes as represented by a clear power ridge along the expected dispersion relations (red points) in the spectra for $3 \le m \le 12$.
In our simulation, these Rossby modes are excited both by the non-axisymmetric random fluctuations in the $\Lambda$-effect and by the non-axisymmetric Lorentz-force, unlike the rotating convection simulation of \citet[][]{bekki2021b} where they are excited by turbulent convective motions alone.
It is implied that our code can be used to study the magnetic cycle dependence of the Rossby modes (or inertial modes in general) in the future.


\section{Summary and Discussions} \label{sec:summary}

In this paper, we have developed a new Babcock-Leighton flux-transport dynamo model of the Sun.
In our model, we do not solve the small-scale convection and focus on the large-scale flows and magnetic structures in a full spherical shell.
The solar-like large-scale mean flows are driven by proper parameterization of the $\Lambda$-effect.
The model operates in a 3D non-kinematic regime, and therefore, is more realistic than the 2D non-kinematic models \citep[e.g.,][]{rempel2006,ichimura2017} and 3D kinematic models \citep[e.g.,][]{yeates2013,miesch2014}.

To better illustrate the major differences from the conventional 2D non-kinematic models and the 3D kinematic models, 
we first carry out a set of simulations for a single BMR with different initial subsurface structure.
We find that, when the BMR has a shallow subsurface structure and a large longitudinal separation, the post-emergence evolution of the BMR becomes significantly changed from those from the conventional models:
Even if the initial BMR is perfectly east-west aligned (zero tilt angle), it begins to acquire a negative tilt angle (which is opposite to the Joy's law).
The strength of the negative tilt angle decreases as the model bipole is embedded deeper in the solar convection zone.
Furthermore, we find a strong asymmetry in the field strengths between the leading and following polarity regions.
The leading polarity field becomes stronger whereas that of the following spot becomes weaker, which is similar to the observations.
These results can be explained by the Coriolis force acting on the flows driven by the Lorentz force of the BMR (see Fig.~\ref{fig:schematic}).

We also carry out the cyclic solar dynamo simulation using the source term of the Babcock-Leighton $\alpha$-effect which is implemented in a 3D manner where the Joy's law tilts are explicitly given.
We have successfully demonstrated that many observational features are reproduced in our model such as the activity cycles with decadal periods, the equatorward migration of the sunspot groups (BMRs), and the poleward transport of the surface radial fields.
The nonlinear saturation of the dynamo occurs due to a strong Lorentz force feedback: 
The magnetic energy of the toroidal field amplified by the $\Omega$-effect is found to exceed the kinetic energy of the differential rotation. 
This strong Lorentz force feedback can be seen in the cyclic modulations of the differential rotation (torsional oscillations) and the meridional circulation.
Note, however, that our study does not exclude other nonlinear dynamo saturation mechanisms such as variability in the Babcock-Leighton process \citep{weber2011,karak2017} and magnetic quenching of the turbulent transport processes \citep[][]{kitchatinov1994,cattaneo1996,yousef2003}.

Since our model is highly sensitive to various model parameters, there are still several disagreements with the solar observations such as a slightly shorter cycle period of $9$ year, stronger radial field strengths at the surface of typical amplitudes of about $200-300$ G, and slightly larger torsional oscillations.
Obviously, the model parameters associated with the subsurface structure of the newly-emerged BMRs will be highly influential, as expected from the discussion in \S~\ref{sec:paramstudy}. 
The magnetic diffusivity $\eta$ also has a substantial impact on the dynamo cycle properties by regulating the diffusive transport of magnetic fluxes in the Sun.
The other important parameters would be $\Lambda_{0}$ that determines the amplitudes of the differential rotation and meridional circulation, and $a_{0}$ that determines the field strengths of BMRs at the surface.
A detailed parameter study is required in the future.

Due to the 3D non-kinematic nature of our model, we find the substantial non-axisymmetric flows that are driven by the Lorentz-force of the BMRs.
These flows have a spatial extent of about 10$^{\circ}$, similar to what is seen in observations \citep[][]{gizon2001,loeptien2017}.
Such flows have been shown to affect not only the evolution of the associated active regions but also the global magnetic field configuration through interaction with other nearby active regions \citep[][]{martinbelda2016}. 
This nonlinear interaction is sensitive to the details of the surface flows which are not yet consistent with the solar observations.

An important physical ingredient still missing in the present model is the enhanced radiative cooling associated with the BMRs, which will affect both the short-term post-emergence evolution of the BMRs and the long-term cyclic dynamo behaviors.
This radiative loss will substantially affect the surface horizontal motions by geostrophycally inducing inflows around the active regions \citep[e.g.,][]{gizon2001,gizon2008}.
These active region inflows are expected to affect the tilt angle \citep[][]{martinbelda2016}, regulate the poleward transport of the poloidal fluxes and limit the buildup of the polar fields \citep{jiang2010}, and thus affect the cycle amplitudes in the Babcock-Leighton solar dynamo \citep{cameron2012}.
Furthermore, it is often argued that the low-latitude branches of the torsional oscillation are attributed to the thermally-induced flows due to the enhanced surface cooling of the BMRs (thermal forcing) \citep{spruit2003,rempel2006,rempel2007}.
In the future model, we plan to include this effect to study how the post-emergence of the BMRs and the nonlinear saturation of the dynamo change in the 3D non-kinematic regime.

Lastly, we note that our code can be used to examine the impact of magnetic fields on various kinds of inertial modes which we found to exist in our simulation (see Fig.~\ref{fig:rossby}).
Recent observations suggest that the amplitudes and frequencies of some of the solar equatorial Rossby modes exhibit a cycle dependence \citep{liang2019}.
If we properly understand the effects of deep-seated magnetic fields on the mode frequencies and eigenfunctions of the equatorial Rossby modes, observations could potentially be used to infer the location and strength of the magnetic fields hidden in the Sun.

\begin{acknowledgements}
We thank an anonymous referee for constructive comments.
We also thank B. Karak for helpful comments on the initial manuscript.
Y. B. was enrolled in the International Max-Planck Research School for Solar System Science at the University of G\"ottingen (IMPRS). 
Y .B. also acknowledges a support from a long-term scholarship program for degree-seeking graduate students abroad from the Japan Student Services Organization (JASSO).
We acknowledge a support from ERC Synergy Grant WHOLE SUN 810218.
All the numerical computations were performed at the Max-Planck supercomputer RZG in Garching. 
\end{acknowledgements}

\bibliographystyle{aa} 
\bibliography{ref} 

\clearpage
\newpage

\begin{appendix} 

\section{Lorentz force of a model BMR} \label{app:A}

\begin{figure*}[ht]
\begin{center}
\includegraphics[width=\linewidth]{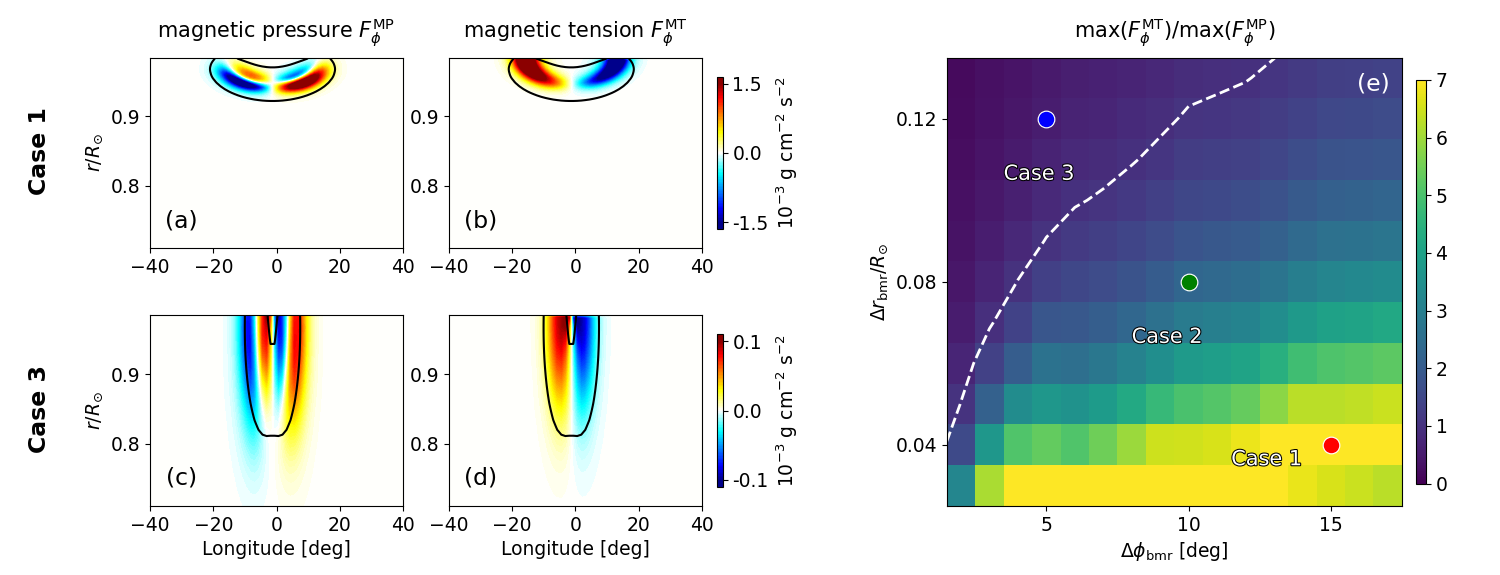}
\caption{
Longitudinal component of the magnetic pressure and magnetic tension forces $F^{\mathrm{MP}}_{\phi}$ and $F^{\mathrm{MT}}_{\phi}$.
(a,b) $F^{\mathrm{MP}}_{\phi}$ and $F^{\mathrm{MT}}_{\phi}$ for the initial BMR for the Case 1, respectively.
The cross sections at $\theta=\theta^{*}$ are shown.
(c,d) The same as panels (a) and (b) for the Case 3.
(e) Ratio between the maximum amplitudes of $F^{\mathrm{MP}}_{\phi}$ and $F^{\mathrm{MT}}_{\phi}$ at the surface ($r=0.985R_{\odot}$) with different combinations of the model parameters $\Delta r_{\mathrm{bmr}}$ and $\Delta \phi_{\mathrm{bmr}}$.
The parameters used in Cases 1--3 are denoted by red, green, and blue circles.
The white dotted line represents the contour line of unity.
}
\label{fig:Fphi_BMR}
\end{center}
\end{figure*}
\begin{figure*}
\begin{center}
\includegraphics[width=\linewidth]{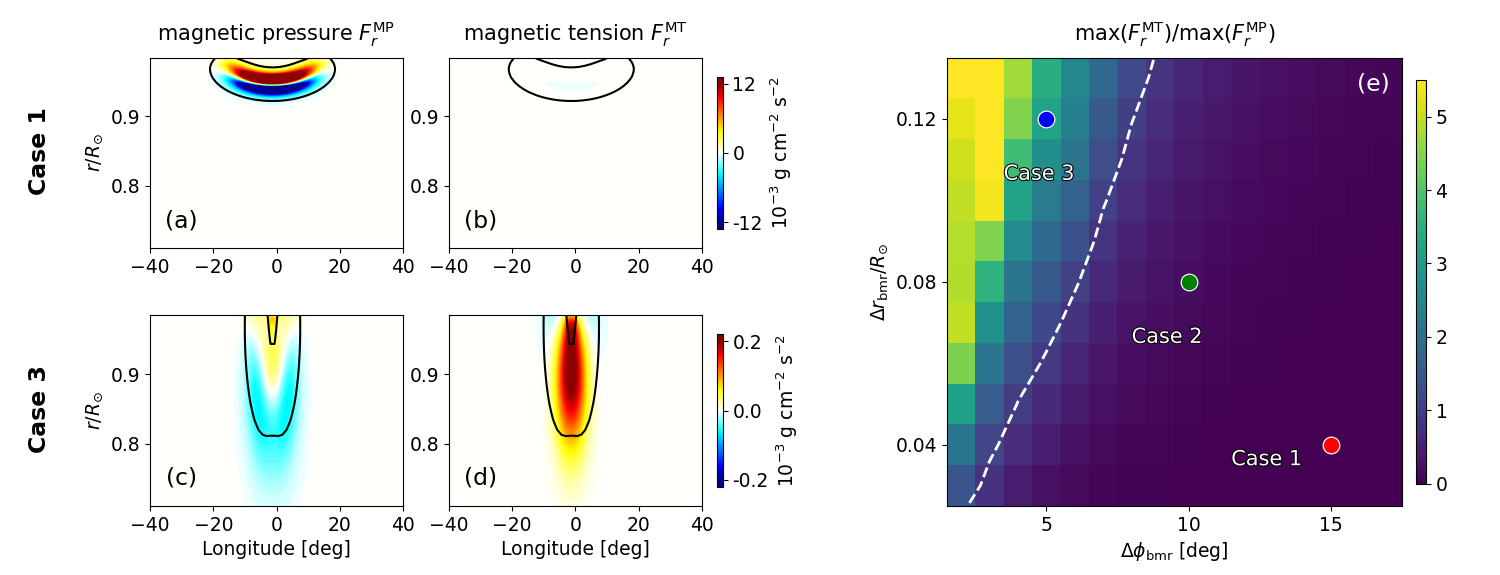}
\caption{
The same as Fig.~\ref{fig:Fphi_BMR} but for the radial component of the magnetic pressure and magnetic tension forces $F^{\mathrm{MP}}_{r}$ and $F^{\mathrm{MT}}_{r}$.
(e) Ratio between the maximum amplitudes of $F^{\mathrm{MP}}_{r}$ and $F^{\mathrm{MT}}_{r}$ at the polarity-inversion line ($\phi=0$).
}
\label{fig:Fr_BMR}
\end{center}
\end{figure*}

In \S~\ref{sec:paramstudy}, it is shown that the post-emergence evolution of a model BMR is very sensitive to its subsurface shape and this is because of the change in the Lorentz force of the BMR.
In this appendix, we show the radial and longitudinal components of the Lorentz force associated with the model BMR discussed in \S~\ref{sec:paramstudy}.
The initial magnetic field is specified by the Eqs.~(\ref{eq:Bic1})--(\ref{eq:Bic3}).
We decompose the Lorentz force per unit mass into the magnetic pressure force ($\bm{F}^{\mathrm{MP}}$) and the magnetic tension force ($\bm{F}^{\mathrm{MT}}$) as
\begin{eqnarray}
    && \frac{1}{4\pi}(\nabla\times\bm{B})\times\bm{B}
    =\underbrace{-\nabla\left( \frac{\bm{B}^{2}}{8\pi}\right)}_{\bm{F}^{\mathrm{MP}}} 
    + \underbrace{\frac{1}{4\pi}(\bm{B}\cdot \nabla)\bm{B}}_{\bm{F}^{\mathrm{MT}}} .
\end{eqnarray}

Figures~\ref{fig:Fphi_BMR}a--d show the longitudinal component of the magnetic pressure and magnetic tension forces for Case 1 and Case 3, respectively.
In Case 1 where the BMR is localized in a shallow surface layer and the two spots are distantly separated in longitude, the longitudinal Lorentz force at the surface is dominated by the magnetic tension force $F^{\mathrm{MT}}_{\phi}$, which acts to pull the two spots together.
On the other hand, in Case 3 where the BMR is anchored deeper in the convection zone and the two spots are located close by each other, the longitudinal component of the magnetic tension force significantly decreases because the field lines are no longer curved at the surface.
Figure~\ref{fig:Fphi_BMR}e shows the ratio between $F^{\mathrm{MT}}_{\phi}$ and $F^{\mathrm{MP}}_{\phi}$ at the surface, implying that the longitudinal Lorentz force is dominated by the magnetic tension force when $\Delta r_{\mathrm{bmr}}$ is small, i.e., $\Delta r_{\mathrm{bmr}} \lesssim 0.05R_{\odot}$.
Within this parameter regime, the BMR is expected to gain a negative (anti-Joy's law) tilt owing to the strong tension-force-driven longitudinal converging flows and the subsequent Coriolis force acting on them.

The radial component of the magnetic pressure and tension forces are shown in Fig.~\ref{fig:Fr_BMR}.
It is seen that the radial Lorentz force is largely dominated by the magnetic pressure force $F^{\mathrm{MP}}_{r}$ in Case 1, by which the plasma inside the flux tube is pushed outward (both radially downward and upward).
The associated pressure disturbances are propagated as (magneto-)acoustic waves, which can be seen in Figs.~\ref{fig:results-case1}a--\ref{fig:results-case2}a.
This initial relaxation occurs on timescales shorter than the dynamical timescale of the BMR evolution.
On the other hand, in Case 3, the radial Lorentz force is dominated by the magnetic tension force $F^{\mathrm{MT}}_{r}$ which is directing upward.
This upward motion, coupled with the Coriolis effect, is expected to induce the retrograde plasma flows inside the flux tube and produce the field asymmetry between the leading and the following spots in Cases 2 and 3, as illustrated in Fig.~\ref{fig:schematic}b.

\section{Cyclic dynamo simulations with different model parameters} \label{app:B}

In \S~\ref{sec:results}, we describe the results from our reference simulation.
In this appendix, we report two additional cyclic dynamo simulations with different model parameters.

\begin{figure*}
\begin{center}
\includegraphics[width=0.7\linewidth]{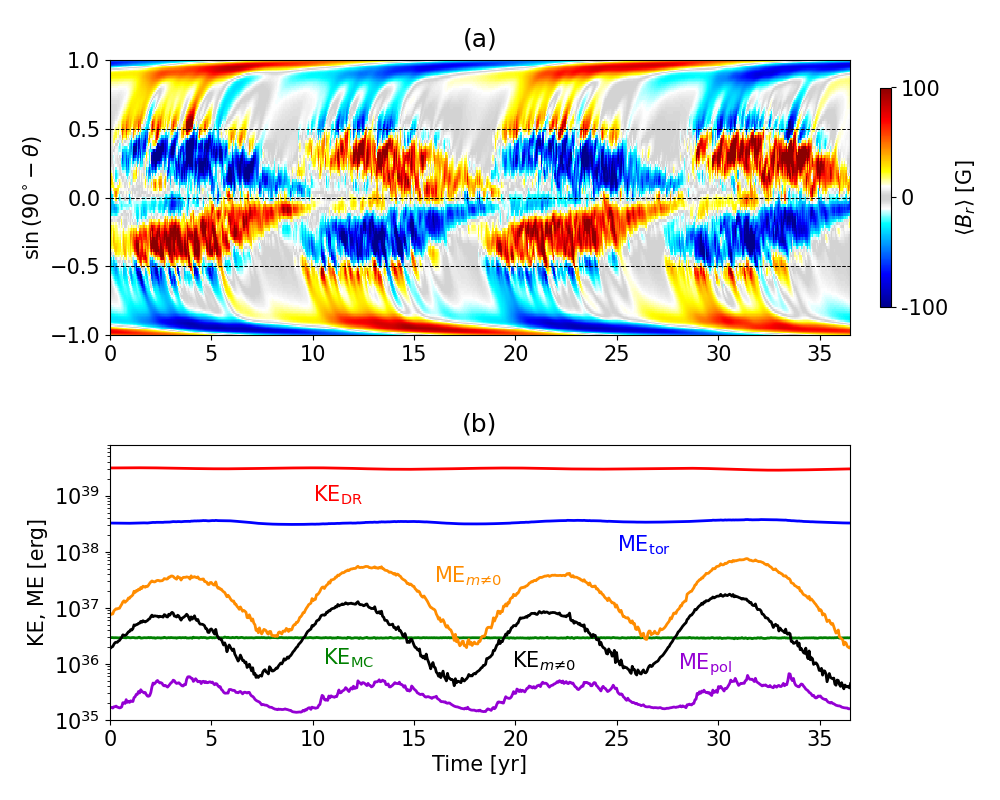}
\caption{
Temporal evolution of the cyclic dynamo simulation with weak Babcock-Leighton $\alpha$-effect ($a_{0}=50$ km s$^{-1}$).
(a) Magnetic butterfly diagram at the surface $r=0.985R_{\odot}$.
(b) Temporal evolution of the volume-integrated kinetic and magnetic energies.
The same definition of colors as in Fig.~\ref{fig:Etime} is used.
}
\label{fig:bfly_weak}
\end{center}
\end{figure*}
\begin{figure*}
\begin{center}
\includegraphics[width=\linewidth]{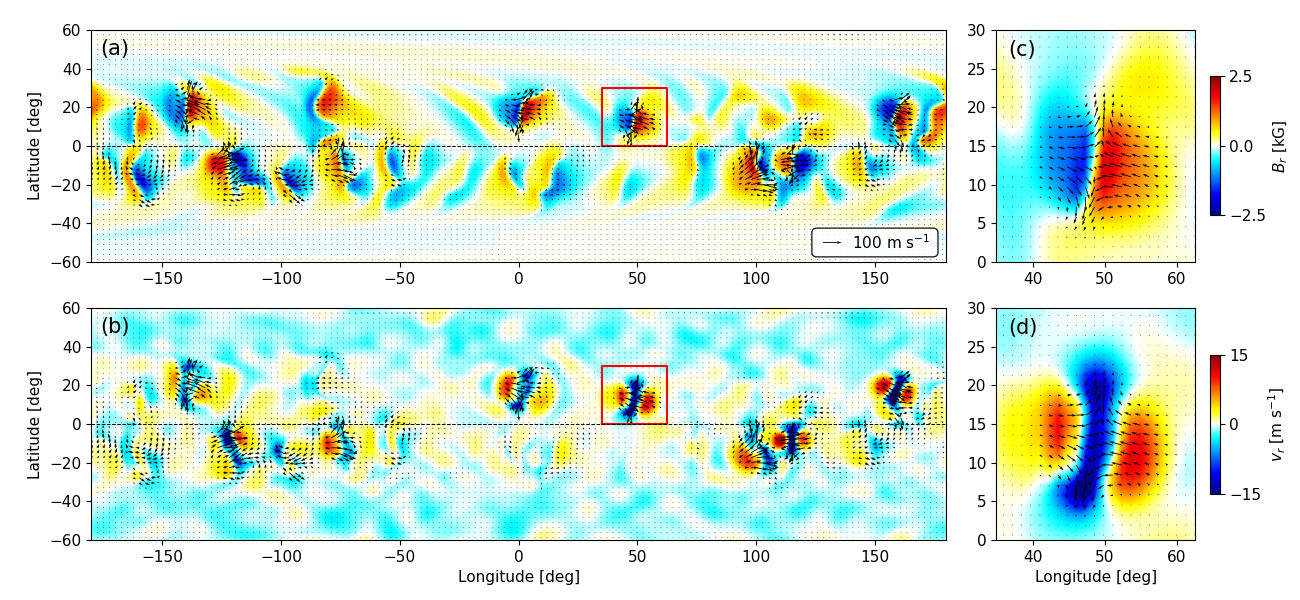}
\caption{
The same as Fig.~\ref{fig:sfvec} but for the dynamo simulation with weak Babcock-Leighton $\alpha$-effect source term ($a_{0}=50$ km s$^{-1}$).
}
\label{fig:surface_weak}
\end{center}
\end{figure*}

\subsection{$a_{0}=50$ km~s$^{-1}$} \label{app:B1}

We carry out the simulation with weaker Babcock-Leighton $\alpha$-effect source, in which $a_{0}$ is decreased to 50 km~s$^{-1}$ while all the other parameters remain unchanged from the reference calculation reported in \S~\ref{sec:results}.
Figure~\ref{fig:bfly_weak}a shows the magnetic butterfly diagram at the surface.
The amplitude of the longitudinally-averaged radial field $\langle B_{r} \rangle$ can be reduced by a factor of $3$ (see Fig.~\ref{fig:bfly}a).
It is seen that the simulation can still reproduce many observed properties of the solar dynamo such as the decadal polarity reversals and the equatorward migration of the activity belt.
We find that the Lorentz force feedback from the dynamo generated fields to the flows is substantially weaker in this simulation.
Figure~\ref{fig:bfly_weak}b shows the volume-integrated kinetic and magnetic energies as functions of time.
It is shown the the kinetic energy of the differential rotation $\mathrm{KE_{DR}}$ is always greater than the toroidal magnetic energy $\mathrm{ME_{tor}}$ by about one order magnitude, suggesting that the $\Omega$-effect operates as if in the kinematic regime without being quenched by Lorentz force feedback.
This is in clear contrast to the reference case shown in Fig.~\ref{fig:Etime}.
In fact, the typical torsional oscillation amplitude is found to be about $1-2$ nHz, which is smaller than those of our reference model and the solar observations.

Figure~\ref{fig:surface_weak} shows the temporal snapshot of the radial magnetic field $B_{r}$ and the non-axisymmetric flows at the surface.
The latitudinal elongation of the BMRs (which is commonly seen in our reference calculation but not observed on the Sun) can be significantly alleviated.

We find that, in the simulation with weak Babcock-Leighton $\alpha$-effect source, some inertial modes exhibit a very strong dependence on the magnetic activity cycles.
This, as well as the other dynamo properties of this simulation, will be reported and discussed in great detail in a future publication (Bekki et al. in prep).

\subsection{$\Delta r_{\mathrm{bmr}}=0.12R_{\odot}$} \label{app:B2}

\begin{figure*}
\begin{center}
\includegraphics[width=\linewidth]{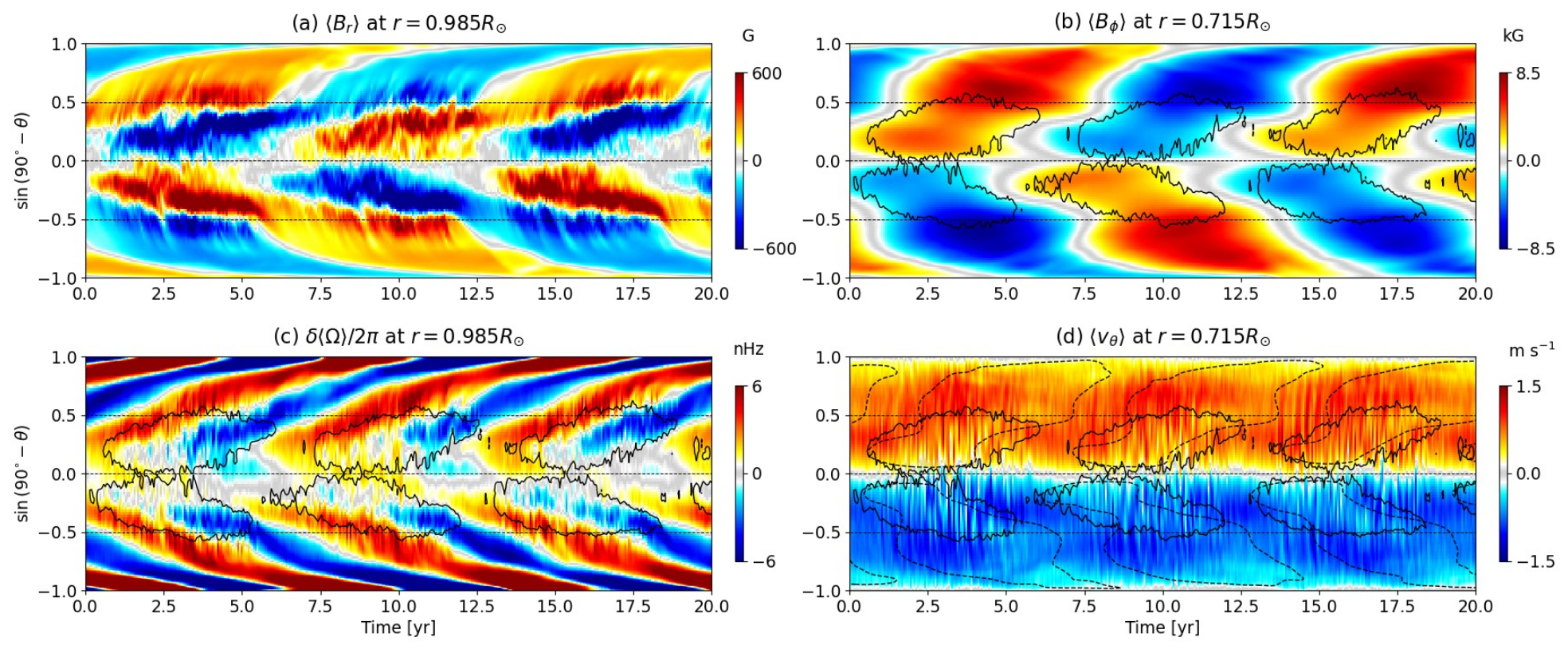}
\caption{
Temporal evolution of the longitudinally-averaged magnetic fields and horizontal velocities for the dynamo simulation with deeper Babcock-Leighton $\alpha$-effect ($\Delta r_{\mathrm{bmr}}=0.12R_{\odot}$).
(a) Azimuthally-averaged radial field $\langle {B}_{r} \rangle$ at the surface $r=0.985R_{\odot}$.
(b) Azimuthally-averaged toroidal field $\langle {B}_{\phi} \rangle$ near the base of the convection zone $r=0.715R_{\odot}$.
Black solid lines are the contours of the emerged BMRs at each time.
(c) Torsional oscillation pattern $\delta \langle \Omega \rangle$ at the surface.
(d) Azimuthally-averaged latitudinal velocity $\langle {v}_{\theta} \rangle$ near the base of the convection zone.
}
\label{fig:meanfield_dsf12}
\end{center}
\end{figure*}

We further carry out the additional simulation where the newly-emerged BMRs have a much deeper radial extent, i.e., $\Delta r_{\mathrm{bmr}}=0.12R_{\odot}$.
All the other model parameters are unchanged from the reference simulation.
Figures~\ref{fig:meanfield_dsf12}a and b show the time-latitude plots of the longitudinally-averaged radial magnetic field $\langle B_{r}\rangle $ at the surface and the toroidal field $\langle B_{\phi}\rangle $ at the base of the convection zone, respectively.
It is shown that, when the BMRs are anchored deep in the convection zone, the emergence latitude of the BMRs tends to propagate poleward in time, in striking contrast to the solar observations.
The cycle period is about 6-7 years, which is shorter than that of the reference simulation and in the observation.
The torsional oscillation $\delta\langle \Omega \rangle$ at the surface also reflects this poleward migration of activity belt, forming clear poleward high-latitude branches, as shown in Fig.~\ref{fig:meanfield_dsf12}c.
We confirm that the longitudinally-averaged meridional flow $\langle v_{\theta}\rangle$ is always equatorward near the base of the convection zone and does not flip the sign via the Lorentz force during the activity cycles (Fig.~\ref{fig:meanfield_dsf12}d).
Therefore, the poleward migration of the activity belt cannot be attributed to the advection by the poleward meridional flow.

We attribute the origin of the poleward migration of activity belt to the $\alpha\Omega$ dynamo waves.
When $\Delta r_{\mathrm{bmr}}$ is increased, the mean poloidal field generation by the Babcock-Leighton $\alpha$-effect (which occurs in response to the toroidal field at the base of the convection zone) is no longer confined in a thin surface layer.
If the locations of the $\alpha$-effect and the $\Omega$-effect are not spatially separated, the dynamo waves cannot be avoided.
In the simulation reported here, the Babcock-Leighton $\alpha$-effect has an effective $\alpha_{\phi\phi}$ which is positive (negative) in the northern (southern) hemisphere.
According to the Parker-Yoshimura sign rule \citep[][]{parker1955,yoshimura1975}, the propagation direction of the $\alpha\Omega$ dynamo wave $\bm{s}_{\alpha\Omega}$ is given by
\begin{eqnarray}
    && \bm{s}_{\alpha\Omega} \propto \alpha_{\phi\phi} \nabla \langle \Omega \rangle \times \bm{e}_{\phi}.
\end{eqnarray}
Therefore, the dynamo waves propagate poleward in low to middle latitudes where $\partial \langle \Omega \rangle /dr$ is positive in the convection zone.
We must note that the $\alpha\Omega$ dynamo waves seen in our simulation are distinct from the conventional ones where the $\alpha$-effect represents the small-scale helical turbulence \citep[e.g.,][]{parker1955,moffatt1978}.

\begin{figure*}
\begin{center}
\includegraphics[width=\linewidth]{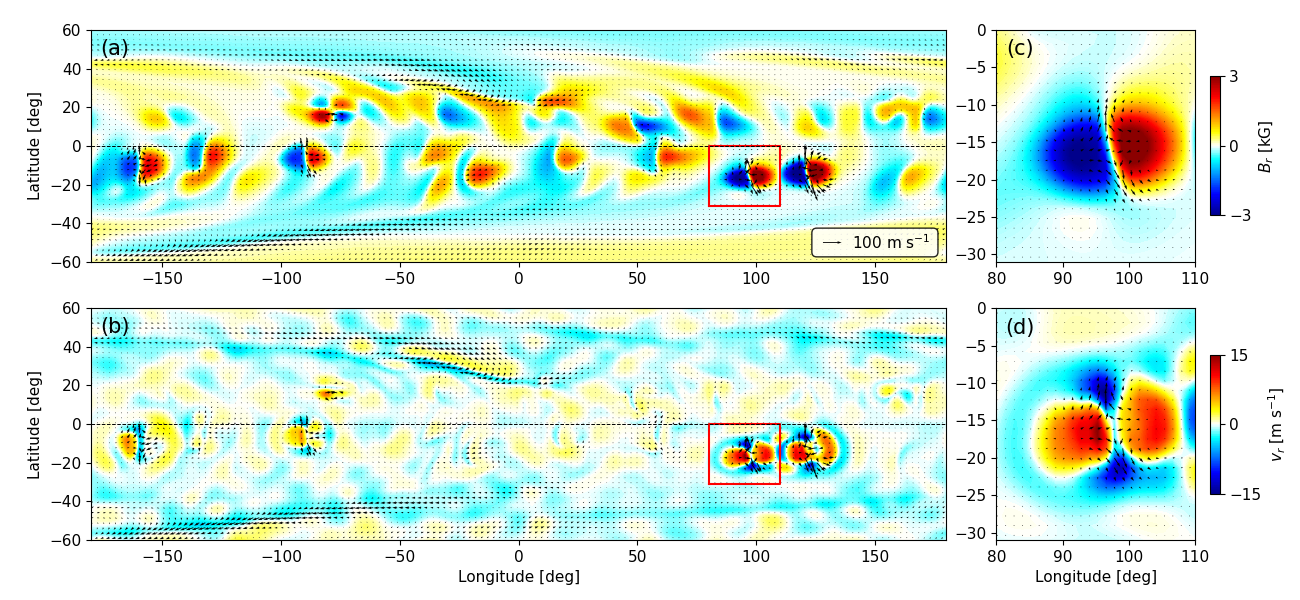}
\caption{
The same as Fig.~\ref{fig:sfvec} but for the dynamo simulation with deeper Babcock-Leighton $\alpha$-effect ($\Delta r_{\mathrm{bmr}}=0.12R_{\odot}$).
}
\label{fig:surface_dsf12}
\end{center}
\end{figure*}

Figure~\ref{fig:surface_dsf12} shows the temporal snapshot of the radial magnetic field $B_{r}$ and the non-axisymmetric flows at the surface from the simulation with deep Babcock-Leighton $\alpha$-effect ($\Delta r_{\mathrm{bmr}}=0.12R_{\odot}$).
Is is seen that the latitudinal elongation of the BMRs which is characteristic in the model with shallow BMRs (see Fig.~\ref{fig:sfvec}) can be significantly relaxed.


\end{appendix}

\end{document}